\documentclass[lettersize,journal]{IEEEtran}
\usepackage{amsmath,amsfonts}
\usepackage{algorithmic}
\usepackage{algorithm}
\usepackage{graphicx}
\usepackage{array}
\usepackage[caption=false,font=normalsize,labelfont=sf,textfont=sf]{subfig}
\usepackage{textcomp}
\usepackage{stfloats}
\usepackage{url}
\usepackage{verbatim}
\usepackage{graphicx}
\usepackage{cite}
\usepackage{multirow}
\usepackage{tabularx}
\usepackage{bm}
\usepackage{enumerate}
\usepackage{booktabs}
\usepackage{tabularx}
\usepackage{caption}
\usepackage[table]{xcolor}

\usepackage{hyperref}
\usepackage{makecell}
\definecolor{lightgray}{rgb}{0.8, 0.8, 0.8}
\usepackage[justification=centering]{caption}

\begin{document}

\title{ZMM-TTS: Zero-shot Multilingual and Multispeaker Speech Synthesis Conditioned on Self-supervised Discrete Speech Representations}
\author{Cheng Gong,~\IEEEmembership{Student Member,~IEEE,} 
Xin Wang,~\IEEEmembership{Member,~IEEE,} Erica Cooper,~\IEEEmembership{Member,~IEEE,} \\
Dan Wells,~\IEEEmembership{Student Member,~IEEE,}
Longbiao Wang,~\IEEEmembership{Member,~IEEE,} Jianwu Dang,~\IEEEmembership{Member,~IEEE,} \\
Korin Richmond,~\IEEEmembership{Senior Member,~IEEE,}
and 
Junichi Yamagishi, ~\IEEEmembership{Senior Member,~IEEE}
\thanks{Cheng Gong, Longbiao Wang and Jianwu Dang are with the Tianjin University, Tianjin, China. Xin Wang, Erica Cooper, and Junichi Yamagishi are with the National Institute of Informatics (NII), Tokyo, Japan. 
Dan Wells and Korin Richmond are with the Centre for Speech Technology Research, University of Edinburgh, United Kingdom.
This work was done when Cheng Gong was a visiting Ph.D.
student at NII. 
(\textit{Corresponding authors: Longbiao Wang; Junichi Yamagishi.})

This work was supported in part by the National Natural Science Foundation of China under Grant (U23B2053, 62176182), the China Scholarship Council (CSC) No. 202206250146, MEXT KAKENHI Grants (21H04906, 21K17775, 21K11951), and the National Research Council of Canada's Ideation Fund: `Small teams – Big Ideas'.
}}

\markboth{Journal of \LaTeX\ Class Files,~Vol.~14, No.~8, Nov~2023}%
{Shell \MakeLowercase{\textit{et al.}}: A Sample Article Using IEEEtran.cls for IEEE Journals}

\maketitle
\begin{abstract}
Neural text-to-speech (TTS) has achieved human-like synthetic speech for single-speaker, single-language synthesis. Multilingual TTS systems are limited to resource-rich languages due to the lack of large paired text and studio-quality audio data. TTS systems are typically built using a single speaker's voice, but there is growing interest in developing systems that can synthesize voices for new speakers using only a few seconds of their speech. This paper presents ZMM-TTS, a multilingual and multispeaker framework utilizing quantized latent speech representations from a large-scale, pre-trained, self-supervised model.
\textcolor{black}{Our paper combines text-based and speech-based self-supervised learning models for multilingual speech synthesis. Our proposed model has zero-shot generalization ability not only for unseen speakers but also for unseen languages.}
We have conducted comprehensive subjective and objective evaluations through a series of experiments.
Our model has proven effective in terms of speech naturalness and similarity for both seen and unseen speakers in six high-resource languages.  
We also tested the efficiency of our method on two hypothetically low-resource languages. The results are promising, indicating that our proposed approach can synthesize audio that is intelligible and has a high degree of similarity to the target speaker's voice, even without any training data for the new, unseen language.
\end{abstract}

\section{INTRODUCTION}
\label{sec:intro}
Text-to-speech (TTS) technology, which converts a text string into a speech waveform\cite{DBLP:conf/interspeech/WangSSWWJYXCBLA17,DBLP:conf/iclr/0006H0QZZL21,DBLP:conf/icassp/ShenPWSJYCZWRSA18,tan2022naturalspeech}, has received significant advancements driven by deep learning.
Advances in this field have been beneficial for a variety of uses including audio-book narration, news readers, conversational assistants and engaging user experiences in virtual worlds.
However, this success is dependent on a rich resource of high-quality data, and thus data requirements have become a bottleneck of both research and the implementation of neural TTS\cite{saeki2023learning}.
Preparing data resources and building TTS systems for a specific language can be costly, let alone repeating this for thousands of languages around the world. 
In this paper, our aim is to build a unified model for a multilingual and multispeaker system, opening possibilities for unseen speakers and unseen language adaptations in low-resource scenarios.

The expansion of the  TTS systems' coverage of the world's languages has attracted significant attention from both academia and industry\cite{DBLP:conf/interspeech/ZhangWZWCSJRR19,DBLP:conf/interspeech/NekvindaD20,DBLP:conf/icml/CasanovaWSJGP22}.
Phonetically-based speech processing systems often require pronunciation dictionaries to map words to a sequence of phonetic units. Recent studies have demonstrated that pre-trained models for phonetic representations, such as PnG BERT \cite{jia2021png}, Mixed-Phoneme BERT \cite{zhang2022mixed} and Phoneme-level BERT \cite{li2023phoneme}, can be beneficial for advanced TTS systems. Motivated by the impressive cross-lingual transferability of multilingual language models, several works \cite{saeki2023learning,xphonebert} perform masked-language model (MLM) pre-training with multilingual text-only data for TTS.
Considering the abundant resources of textual data in the real world, which are usually more readily accessible compared with audio, it is worthwhile to investigate the utilization of language models pre-trained on large text resources for phoneme representations in multilingual synthesis tasks.

To synthesize speech with a target speaker's timbre, end-to-end models usually learn a speaker embedding during training that enables speaker selection at inference time.
Furthermore, several of these models enable zero-shot speaker synthesis
using a speaker vector extracted from a short audio clip\cite{NEURIPS2018_6832a7b2,9054535}. 
Despite recent advances, the similarity gap between seen and unseen speakers remains an open research question, and training these models still requires a significant number of speakers, posing challenges in developing high-quality models for low-resource languages\cite{DBLP:conf/icml/CasanovaWSJGP22}.
Furthermore, in real-world data, most of the attributes
that represent speech are interrelated and difficult to disentangle.
To control a multilingual and multispeaker system, it is crucial to disentangle speaker and language\cite{DBLP:conf/interspeech/ZhangWZWCSJRR19}.
Efforts using adversarial learning or consistency loss \cite{DBLP:conf/interspeech/ZhangWZWCSJRR19,DBLP:conf/icml/CasanovaWSJGP22} have been made to mitigate the performance degradation resulting from this entanglement. 

However, these state-of-the-art multilingual and multispeaker systems mainly rely on a large amount of training data, which is not available for every language. The intermediate features used in these studies are often Mel spectrograms, which have a high correlation in time and frequency, making it difficult to disentangle speaker-dependent information.
Fortunately, self-supervised learning (SSL) speech representations \cite{DBLP:conf/nips/BaevskiZMA20,hsu2021hubert,9814838} have been shown to be useful for speech processing tasks, such as speech recognition \cite{9746223,chen2022large}, speech reconstruction \cite{NEURIPS2021_87682805}, and voice conversion \cite{9746430}. 
The learned representations are used as input for a supervised model, which is often fine-tuned to improve task performance or reduce the labeled data required. Recently, several TTS models\cite{wang2023neural, liu23d_interspeech} have started using discrete vector-quantized speech representations as intermediate features instead of traditional Mel spectrograms for prediction.
As a result, the quantized output has less speaker-dependent information than the Mel spectrograms\cite{liu23d_interspeech}. While a designed fine-tuning protocol for the learned representations indeed improves automatic speech recognition (ASR) performance \cite{9688093}, in low-resource settings, relatively little attention has been paid to the study of TTS models using learned SSL representations.

\textcolor{black}{Recently, large-scale TTS systems \cite{wang2023neural,borsos2023audiolm,kharitonov2023speak,le2024voicebox} that leverage data-driven representations, i.e., either discrete tokens or continuous vectors from an auto-encoder \cite{zeghidour2021soundstream,defossez2023high} have been widely adopted for zero-shot speech synthesis. 
Several of these models \cite{wang2023neural,borsos2023audiolm} employ an autoregressive architecture to generate discrete tokens one by one, like a language model. These autoregressive models suffer from a slow inference speed, unstable prosody, and word skipping/repeating issues. To overcome this limitation, non-autoregressive large-scale TTS systems, such as NaturalSpeech 2/3 \cite{shen2023naturalspeech,ju2024naturalspeech} and HierSpeech++ \cite{lee2023hierspeech++}, have been investigated.
However, all the aforementioned large-scale TTS systems heavily rely on abundant data and primarily focus on resource-rich languages rather than low-resource ones.}

In this paper, we propose a unified zero-shot multilingual multispeaker TTS framework called ZMM-TTS that leverages discrete speech representations from a multilingual speech-based SSL model.
ZMM-TTS consists of a text-to-discrete speech representations module (txt2vec) and a representations-to-waveform module (vec2wav).
Regarding the txt2vec module, we adopt a standard end-to-end TTS architecture comprising token embeddings, an encoder, and a decoder to predict discrete representations.
To utilize the strong cross-lingual transferability of multilingual language models, we adopt a pre-trained large-scale multilingual language model for phoneme representations.
For the vec2wav module, an additional multi-stage and multi-head vector quantization (VQ) model was adopted to preserve the discrete information at different time resolutions.

In summary, we have made the following contributions:
\begin{itemize}
\item 
We propose ZMM-TTS, a multilingual and multispeaker framework using discrete audio representations from a large-scale multilingual pre-trained self-supervised model as an intermediate representation to replace the widely-used Mel spectrograms.
\item
We investigate the impact of various input representations on multilingual synthesis tasks that use SSL discrete representations. \textcolor{black}{Our research combines phoneme representations from a pre-trained text-based multilingual language model and speech-based SSL representations in a speech synthesis task.}
\item
ZMM-TTS can perform zero-shot multispeaker TTS in a target language with high quality and speaker similarity using only a few seconds of speech during inference.
\item 
We also verify the effectiveness of the ZMM-TTS in low-resource and zero-shot scenarios on two hypothetically low-resource languages.
\textcolor{black}{``Zero-shot'' refers to the capability of synthesizing speech in a low-resource language without training or fine-tuning the model using the data from that language.}
This observation is quite promising, as it indicates that our proposed approach can synthesize intelligible
audio with high speaker similarity, even when no specific training data are available for the previously unseen language. 
Note that ``unseen" in this context refers to the absence of dedicated training data for ZMM-TTS, while the unseen language is still
included in the training data of both the pre-trained language model and self-supervised model.

\end{itemize}
We encourage the reader to listen to our samples, which can be found at \url{https://gongchenghhu.github.io/TASLP-demo/}. \textcolor{black}{The source code has been released on \url{https://github.com/nii-yamagishilab/ZMM-TTS}.}

\textcolor{black}{
The remainder of this paper is organized as follows. 
Section~\ref{sec:related_work} provides a comprehensive review of related work in the field. 
Section~\ref{sec:method} presents a detailed description of the proposed approach. 
The experimental setup is described in Section~\ref{sec:exp_setup}.
In Section~\ref{sec:exp_result_1}, we evaluate and analyze the proposed system using six languages, in particular, the quality of unseen speakers' voices. Section~\ref{sec:exp_2} evaluates and analyzes the proposed system, assuming two unseen low-resource languages. Finally, Section~\ref{sec:conclusion} concludes the paper, summarizing the findings and highlighting future research directions. In Appendix~\ref{sec:appendix}, we show a comparison with the latest large-scale models that utilize larger amounts of data.
}
\section{RELATED WORK}
\label{sec:related_work}
This section reviews related work about zero-shot and multilingual speech synthesis. 
We also review related studies on speech synthesis with the aid of SSL representations, especially in the multilingual cases.
\subsection{Multilingual Speech Synthesis Systems}
\subsubsection{Zero-shot multispeaker TTS}
The concept of zero-shot multispeaker TTS was initially introduced in \cite{arik2018neural} and the main idea is to employ embeddings from an external speaker encoder as a reference signal.
Subsequent research by \cite{9054535} aimed to reduce the quality gap between seen and unseen speakers by utilizing more informative embeddings. \cite{wang2018style} further explored
the utilization of attentive speaker embeddings for the purpose of encoding general speaking styles, while \cite{casanova2021sc} proposed a speaker-conditional architecture and investigated a flow-based decoder capable of generating unseen speaker voices. Despite these efforts, the task of zero-shot multispeaker TTS has yet to be fully solved. Moreover, capturing a wide range of voice properties requires a substantial amount of high-quality data from numerous varied speakers.

Regarding multispeaker multilingual TTS, a primary idea is to train the decoder by incorporating learnable language and speaker embeddings\cite{DBLP:conf/interspeech/ZhangWZWCSJRR19}.
However, the majority of speech attributes are interrelated and difficult to disentangle through embedding tables alone.
Attempts have been made to mitigate the decline in multi-language multispeaker performance caused by speaker and language entanglement, especially in cross-lingual scenarios. \cite{DBLP:conf/interspeech/ZhangWZWCSJRR19} incorporated adversarial domain training, enabling them to transfer different voices between languages. \cite{xin2021disentangled} proposed the use of mutual information minimization to keep speaker consistency in cross-lingual synthesis. \cite{yang2022cross} utilized joint training incorporating a speaker classifier to enhance speaker similarity.

\subsubsection{System structure}
Previous multilingual TTS models \cite{DBLP:conf/interspeech/ZhangWZWCSJRR19,DBLP:conf/interspeech/NekvindaD20} are primarily built using Tacotron \cite{DBLP:conf/interspeech/WangSSWWJYXCBLA17,DBLP:conf/icassp/ShenPWSJYCZWRSA18}.
When synthesizing speech, Tacotron-based models often repeat or skip words due to an autoregressive approach that uses attention to align input text and target speech.
Instead of autoregressive iteration, several non-autoregressive \cite{DBLP:conf/aaai/Li0LZL19,NEURIPS2019_f63f65b5,DBLP:conf/iclr/0006H0QZZL21,DBLP:conf/icassp/Lancucki21} models adopt Transformer encoder and decoder blocks to generate Mel spectrograms in parallel. 
For example, \cite {yang20g_interspeech} create a multilingual speech synthesis system including a Transformer-based acoustic predictor and a WaveNet \cite{DBLP:conf/ssw/OordDZSVGKSK16} neural vocoder.
\cite {badlani23_interspeech} presented a multilingual, multiaccented, multispeaker speech synthesis model based on RADTTS \cite{shih2021rad} which is a parallel flow-based generative model.  
Recently, several fully end-to-end multilingual speech synthesis systems \cite{DBLP:conf/interspeech/ChoJLW22,DBLP:conf/icml/CasanovaWSJGP22} were built on VITS\cite{kim2021conditional}, 
which could produce natural-sounding synthesized speech. Moreover, VITS requires training a single model, so there is no need for an independent vocoder.
Among these fully end-to-end systems, YourTTS \cite{DBLP:conf/icml/CasanovaWSJGP22}, which aims at zero-shot multispeaker TTS and zero-shot voice conversion, presented promising results for a few language combinations.
However, it had limited success in transferring to languages with few speakers and used a curriculum learning approach, making training cumbersome.

Unlike these systems that all use acoustic features such as the Mel spectrograms as intermediate features, we built a multilingual synthesis system that is based on SSL representations to take advantage of better speaker/content disentanglement in discrete SSL representations.
\subsection{Input Representations}
\subsubsection{Characters or Phonemes}
Traditionally, end-to-end TTS models have utilized input representations in the form of characters or phonemes\cite{DBLP:conf/interspeech/WangSSWWJYXCBLA17,DBLP:conf/iclr/0006H0QZZL21}. 
Employing each character as the default input for end-to-end TTS models requires the model to implicitly learn how to convert graphemes to phonemes as part of the synthesis task. 
Incorporating phoneme input simplifies the TTS task, eliminating the need for the model to learn complicated grapheme-to-phoneme rules for languages such as English. 
Furthermore, language-independent pronunciation information can be obtained through language expertise, such as making a unified set of all the languages in a multilingual model, often derived from the International Phonetic Alphabet (IPA) \cite{saeki2023learning,DBLP:conf/interspeech/ChenCLMCWX19}. 
\subsubsection{Pre-trained representations}
Over the past decade, self-supervised pre-training on extensive text corpora, employing language model (LM) or MLM objectives, has demonstrated remarkable success in the field of TTS\cite{jia2021png,zhang2022mixed,li2023phoneme}.
Toward multilingual tasks, \cite{xphonebert} proposed XPhoneBERT which is the first pre-trained multilingual model for phoneme representations for TTS. 
Based on the monolingual experiment results, using XPhoneBERT as a phoneme encoder for input has been shown to significantly enhance the performance of a powerful neural TTS model in terms of naturalness and prosody. 
It also helps to produce high-quality speech using low-resource training data.
\cite {saeki2023learning,10444075} conducted pre-training on a multilingual text-only dataset using an MLM. They then trained this model in a supervised manner on paired text and audio data, while keeping the language-aware embedding layer frozen. This approach enables the model to synthesize languages that are not present in the paired data but exist in the text-only data.

Alternative approaches adopt UTF-8 bytes \cite{DBLP:conf/icassp/LiZSWC19} or phonological features (PFs) \cite{staib20_interspeech,wells2021cross} as input representations.
UTF-8 byte representations encode typographic rather than phonological information. The model does not learn unseen byte combinations, requiring retraining when enrolling in a new language.
For PFs, IPA symbols can be converted to PFs in accordance with certain aspects of their articulation, as implemented in Epitran \cite{Mortensen-et-al:2018}.
Although this conversion is based on an explicit database of IPA symbols and diacritics, it can be considered universally applicable to any language for which IPA transcription is available.
However, there are no pre-trained phoneme-level encoders similar to XPhoneBERT currently available.
The effect of different input representations is still not clear in multilingual speech synthesis with SSL representations. Taking into account the complexity of the model and the support for new languages and low-resource scenarios, we evaluated the effects of using characters, IPA, and XPhoneBERT as input for multilingual TTS. 
\subsection{SSL Representations for speech synthesis}
Well-designed proxy tasks enable SSL to explore the characteristics of unlabeled data and enhance the model's representation ability. The derived representations of the SSL model have useful information for many downstream tasks\cite{wang2022wav2vec,DBLP:conf/interspeech/SiuzdakDRJ22,lin2021s2vc,chen2022does,10129796}. Wav2vec 2.0 \cite{DBLP:conf/nips/BaevskiZMA20} is the most popular pre-trained model that uses SSL on a large amount of unlabeled speech data, and it has demonstrated strong representation abilities in ASR tasks and achieved outstanding results.
A multilingual wav2vec 2.0, denoted as XLSR-53, was also proposed in \cite{DBLP:journals/corr/abs-2006-13979}. 
The results showed a significant improvement over monolingual wav2vec 2.0, particularly in languages with limited resources. 

\textcolor{black}{For SSL in monolingual TTS downstream tasks,} \cite{DBLP:conf/interspeech/SiuzdakDRJ22} introduced WavThruVec, a two-stage architecture that employed high-dimensional wav2vec 2.0 embeddings as an intermediate speech representation.
\textcolor{black}{The paper \cite{du2022vqtts} proposed VQTTS, which uses the vec2wav vocoder to train a non-autoregressive model that maps text to discrete tokens. MQ-TTS \cite{chen2023vector} proposed using multiple codebooks to quantize intermediate features on real-world spontaneous speech.}
Using wav2vec 2.0 acoustic representations of synthesized audio, \cite {10129796} improved prosody modeling by incorporating prosodic characteristics from neighboring utterances.

\textcolor{black}{For multilingual tasks,} \cite{liu23d_interspeech} demonstrated that the VQ features in wav2vec 2.0 contain significantly less speaker information than the Mel spectrograms, and proposed DSE-TTS, a cross-lingual TTS that encodes timbre-related and linguistic-related speaker information separately. In the paper \cite{liu22p_interspeech}, the author proposed extracting unsupervised phonetic representations (UPR) from wav2vec 2.0 for multilingual speech synthesis when target language pronunciation dictionaries are unavailable.
In the work of \cite{wells23_interspeech}, the author built a TTS system for the Scottish Gaelic language based on self-supervised discrete acoustic unit sequences. \textcolor{black}{Scaling TTS to many languages was proposed in \cite{saeki2024extending} by leveraging massively multilingual joint speech and text representation learning.}

However, integrating SSL-based representations into TTS in low-resource settings is still a novel concept, and it is unclear which type of input representation is most suitable for TTS. 
Furthermore, these methods mostly use monolingual speech representations rather than multilingual speech representations, which may benefit low-resource scenarios.
\subsection{Large-scale TTS systems}
Recently, considerable progress has been made in zero-shot TTS by scaling up corpus and model sizes. These large-scale TTS systems \cite{wang2023neural,borsos2023audiolm,kharitonov2023speak,le2024voicebox} usually quantize the continuous speech waveform into discrete tokens through neural audio codec models and model these tokens with autoregressive LMs.
For example, VALL-E \cite{wang2023neural} was the first neural codec language model for speech synthesis utilizing a discrete audio unit and language models. 
In addition to autoregressive architectures, several large-scale models \cite{shen2023naturalspeech,ju2024naturalspeech,lee2022hierspeech,lee2023hierspeech++,li2024styletts} adopt non-autoregressive architectures to improve model generation speed and robustness. While several large-scale models \cite{zhang2023speak,le2024voicebox,barrault2023seamless}, such as VALL-E-X \cite{zhang2023speak}, are multilingual and have been trained on extensive corpora in both Chinese and English, current large-scale models are more focused on generalization to unseen speakers rather than on adapting to unseen languages. These large-scale models perform well only in resource-rich languages while lacking application in low-resource languages.

\textcolor{black}{The differences between our ZMM-TTS system and previous large-scale models are highlighted in Table~\ref{tab:tts_comparison}. 
Compared with previous large-scale models, ZMM-TTS is designed to have the language adaptation ability for low-resource and unseen languages.
Typically, previous models only validated one language or a few high-resource languages. In contrast, the ZMM-TTS model aims to support not only unseen speakers but also unseen languages.}

\begin{table*}[t]
  \setlength\tabcolsep{5pt}
  \caption{\textcolor{black}{Comparing ZMM-TTS with previous large-scale TTS models on task capabilities.}}
  \centering
  \begin{tabular}{llll}
\toprule
          \textbf{Systems} &\textbf{Languages} &\textbf{Few-shot} &\textbf{Zero-shot}  \\ \midrule
\cite{borsos2023audiolm,wang2023neural,ju2024naturalspeech,li2024styletts,lee2022hierspeech,lee2023hierspeech++,kharitonov2023speak}  &Single &Validated only on one language &Unseen speakers \\
 \cite{zhang2023speak,le2024voicebox,barrault2023seamless} &Multiple &Validated only on high-resource languages &Unseen speakers      \\ 
ZMM-TTS  &Multiple  &Adaptable to multiple low-resource languages &Unseen speakers and languages  \\
\bottomrule
\label{tab:tts_comparison}
\end{tabular}
\vspace{-0.4cm}
\end{table*}

\section{METHOD}
\label{sec:method}
Our proposed model is a multilingual and multispeaker synthesis system that uses self-supervised discrete speech representations from a pre-trained multilingual wav2vec 2.0 (XLSR-53) model. 
Quantization in XLSR-53 is based on product quantization by choosing quantized representations from codebooks $\boldsymbol{C}\in\mathbb{R}^{G\times M\times D}$ with $M=320$ entries each, $G=2$ are two separate codebooks in XLSR-53 and each quantization representation has $D=384$ dimensions.

Given a dataset $\mathcal{S} = \{x_i
, y_i\}$, where $y$ is an audio sample and $X = \{x_0, x_1, . . . , x_{T_m}\}$ is its corresponding text sequences, 
we use a pre-trained XLSR-53 model to encode each audio sample into discrete code index sequences $V=\{{v_{1}^{1:G},v_{2}^{1:G},...,v_{T_c}^{1:G}\}}$ and the corresponding discrete representation is $R=\{r_{1}^{1:G},r_{2}^{1:G},...,r_{T_c}^{1:G}\}$.
Obviously, there is a one-to-one correspondence between $V$ and $R$, and once we obtain $V$, we can obtain $R$ by looking up codebooks $\boldsymbol{C}$ when $r=C[:,v,:]$.

\begin{figure}[tbp]
  \centering \includegraphics[width=\linewidth]{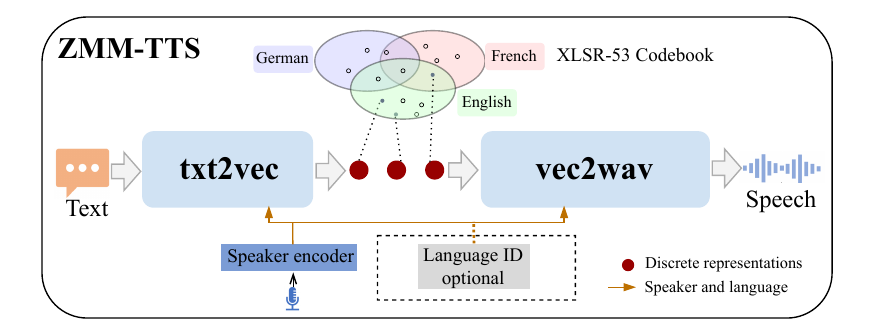}
  \caption{\textcolor{black}{Overview of ZMM-TTS. The modules txt2vec and vec2wav are trained independently. Language ID is utilized for high-resource languages and few-shot adaptation, but it is not used for direct inference without fine-tuning.}}
  \label{fig:1}
  \vspace{-0.4cm}
\end{figure}

Figure \ref{fig:1} depicts the architectures in which our proposed model can be implemented. 
It consists of two steps, as shown in the following formula. 
First, the discrete code index sequences $V$ are predicted from the text $x$ and the codebook of XLSR-53 $\boldsymbol{C}$ is looked up to obtain the discrete representations $R$, and then the waveform is predicted from the discrete representations $R$.
\begin{equation}
\begin{split}
\text{txt2vec}:(x, S, L) \rightarrow R,\\
\text{vec2wav}:(R, S, L) \rightarrow \text{Mel} \rightarrow  \text{Speech} \quad \texttt{OR},\\
(R, S, L) \rightarrow \text{Speech}, \\
\end{split}
\label{eq:1}
\end{equation}
where $S$ denotes the speaker representation and $L$ represents the language identity. 
\textcolor{black}{Note that we discard the language input and related embedding layers, enabling direct inference on unseen languages.}

The txt2vec module is similar to the acoustic model in prior speech synthesis approaches, so we constructed it on the basis of the popular acoustic model FastSpeech. Vec2wav is similar to a vocoder. To obtain a broader understanding of the potential
complementarity and differences between SSL representations and
Mel features, we proposed two distinct approaches to implement vec2wav. The first is to convert discrete representations $R$ into Mel spectrograms and then convert them through an additional vocoder, and the second is to directly convert $R$ into audio. Both methods make use of the up-sampling operation based on transposed convolution in HiFi-GAN \cite{NEURIPS2020_c5d73680}, as in \cite{DBLP:conf/interspeech/SiuzdakDRJ22}.
It should be noted that although there are $G$ different codebooks, in the vec2wav model, the input $R$ is represented by flattened representations from $G$ codebooks together. 
In the following two subsections, we will provide an explanation of the two components, respectively.
\subsection{Predicting Discrete Code Index from Text}
\begin{figure*}[htbp]
  \centering \includegraphics[width=0.9\linewidth]{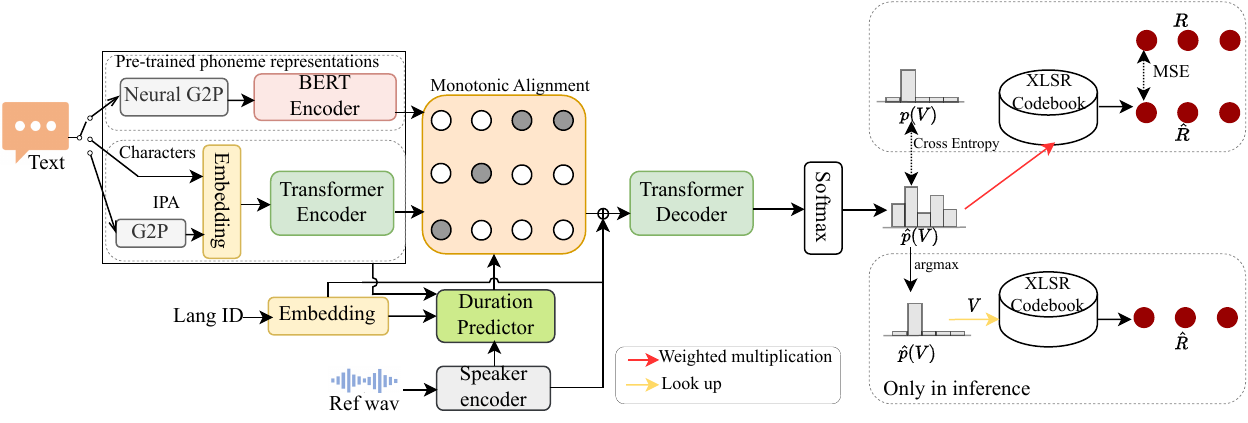}
  \begin{center}
  \caption{Architecture of our proposed txt2vec model.}
  \label{fig:txt2vec}
  \end{center}
  \vspace{-0.8cm}
\end{figure*}
The txt2vec model is based on FastSpeech \cite{NEURIPS2019_f63f65b5}, a non-autoregressive (NAR) acoustic model with explicit duration modeling. 
In our work, we use a learnable aligner \cite{badlani2022one} instead of relying on the predicted phonemes durations of an autoregressive teacher TTS model. This aligner is more robust for long sequences.
The decoder part is the same as that of FastSpeech, which is a Feed-Forward Transformer (FFT). 
For the encoder part, depending on the different types of input representations, it can be either FFT-like or BERT-like. 
Figure \ref{fig:txt2vec} shows the structure of txt2vec.

\subsubsection{Input representations and text encoder}
We attempt to use different input representations including characters, IPA, and pre-trained phoneme representations for multilingual TTS.
\begin{enumerate}[a)]
\item{\emph{Characters:}} 
To extend a character-based input vocabulary to multilingual TTS systems, we only need to concatenate character sets in the training corpus for each language.
\item{\emph{Phonemes-IPA:}} To bring together the many languages under the same input space, we also use IPA, mapping all orthographic text into their IPA representations through the use of Epitran\cite{Mortensen-et-al:2018}.
\item{\emph{Pre-trained phoneme representations:}} 
First, we convert text sentences into a sequence of phonemes using the CharsiuG2P\footnote{https://github.com/lingjzhu/CharsiuG2P} toolkit because of the input labels on which XPhoneBERT relies. 
Then, we convert this sequence of phonemes into a sequence of phoneme representations using a pre-trained XPhoneBERT. 
XPhoneBERT uses the BERT-base architecture, pre-trained on RoBERTa with 330M phoneme-level sentences across 100+ languages.
Here, we use XPhoneBERT as an input phoneme encoder rather than FFT, which is used as the phoneme encoder in FastSpeech.
\end{enumerate}
In summary, when characters or IPA symbols are used as input, FFT is used as the encoder. However, when phonemes from CharsiuG2P are used as input, pre-trained XPhoneBERT is used as the encoder.
\subsubsection{Multispeaker and multilingual control}
To enable control over language and speaker identity, we add language and speaker embeddings as inputs to the decoder and the duration predictor.
To extract speaker embeddings, we use a pre-trained ECAPA-TDNN speaker encoder model.
For language embeddings, we adopt a trainable language embedding table to extract the language embedding of the target language. 
In detail, with the target language ID, a 64-dimensional language embedding can be produced by a language look-up table. 

\subsubsection{Loss Function}
In this
txt2vec stage, the text encoder operates on the input tokens $x$ and produces hidden states, which, combined with language and speaker embeddings, are used to predict the duration $d$ by the duration predictor. 
The up-sampled encoder outputs, together with the language and speaker embeddings, are used as the decoder input.
The decoder then generates discrete values for both $V$ and $R$.

During training, we first adopt a classification task that is optimized by the cross entropy loss between the $\hat{p}(v)$ (predicted probability distributions of $V$) and the target probability distributions $p(v)$.
Given a sequence of decoder output features $Q=\{q_1,q_2,...,q_t\}$, the $p(v)$ can be obtained through 
\begin{equation}
\begin{split}
h_{t}^{g} = \texttt{Split}(F(q_t)),\\
p(v_{t}^{g}|h_{t}^{g})=\texttt{Softmax}(h_{t}^{g}),
\end{split}
\label{eq:3}
\end{equation}
where $F$ is a fully-connected layer that maps the $q_t$ to $G \times D$ dimensions. Subsequently, it further partitions based on the number of codebooks, $G$, to obtain the hidden representation with $D$ dimensions for computing the corresponding classification probability value. 
The classifications loss $\mathcal{L}_{cla}$ can be represented as 
\begin{equation}
\mathcal{L}_{cla} =  \frac{1}{T}\frac{1}{G} \sum_{t=1}^{T}\sum_{g=1}^{G} \texttt{CrossEntropy}(\hat{p}(v_{t}^{g}), p(v_{t}^{g}))
\end{equation}

To learn the cross-lingual information from shared quantized latent speech representations in continuous space, we also incorporated predictions for $R$ during training as
\textcolor{black}{
\begin{equation}
\begin{split}
\hat{r}_{t}^{g}=\sum_{i=1}^{M} p(v_{t}^{g}=i|h_{t}^{g})C_{g,i}
\end{split}
\label{eq:4}
\end{equation}}
\textcolor{black}{Eq. (4) is a weighted sum of the code vectors, where the weights are the predicted probabilities of choosing the codes.
In Eq. (4), $p(v_{t}^{g}=i|h_{t}^{g})$ represents the probability of predicting the code index value $v_{t}^{g}$ as $i$ at time $t$. Here, $g\in G$ denotes the g-th codebook and 
$C_{g,i}$ denotes the vector corresponding to the i-th index in the g-th codebook at $C^{G \times M \times D}$.} 
The regression L2 loss $\mathcal{L}_{MSE}$ between the predicted and targeted $R$ could be represented as
\textcolor{black}{
\begin{equation}
\mathcal{L}_{mse} =  \frac{1}{T}\sum_{t=1}^{T}\frac{1}{G}\sum_{g=1}^{G} {\left \| \hat{r}_{t}^{g} - r_{t}^{g} \right\|}^2_2
\end{equation}}
The total loss consists of the sum of the distance between the ground-truth and predicted discrete features and the duration loss:
\begin{equation}
\mathcal{L}_{total} = \mathcal{L}_{cla} + \mathcal{L}_{mse}+ \mathcal{L}_{dur}
\end{equation}
Given the monotonic alignment assumption between text tokens and discrete representation frames, the $\mathcal{L}_{dur}$ is optimized by maximizing the joint likelihood of each token and frame to find the most likely monotonic path \cite{badlani2022one,shih2021rad}.

Note that since $p(v_{t}^{g})$ is a one-hot vector, the $r_{t}^{g}$ is only one entry in the codebook instead of being multiplied by softmax 
weights as in Eq. (5). Therefore, in inference, we directly select the maximum probability index in the $P(i|h_t^g)$, and look up the corresponding $(\hat{r}_{t}^{g})$ in the code \textcolor{black}{$C^{G \times M \times D}$} as the input of the next-stage model:
\begin{equation}
\begin{aligned}
v_t^g = \arg\max(P(i|h_t^g)) \\
\hat{r_t^g} = C[g,v_t^g,:]
\end{aligned}
\end{equation}

\subsection{Predicting a waveform from discrete representations}
Although most methods convert features directly into audio \cite{DBLP:conf/interspeech/SiuzdakDRJ22}, spectral features may still have a few advantages in several low-resource tasks \cite{berrebbi2022combining}. To obtain a broader understanding of the potential complementarity and differences between SSL representations and Mel features, we adopted two different ways to implement the vec2wav model: with and without an independent Mel-based vocoder as in the following formula,
\begin{equation}
\begin{split}
\text{vec2wav}:(\text{vec2mel} + \text{vocoder}) || (\text{vec2wavVQ (R)})\\
\end{split}
\label{eq:8}
\end{equation}
where vec2mel and vec2wavVQ (R) represent the two methods we propose to transfer discrete representations to the Mel spectrograms and the waveform, respectively.
\subsubsection{vec2wav model with Mel-based vocoder}
\begin{figure}[t]
  \centering \includegraphics[width=\linewidth]{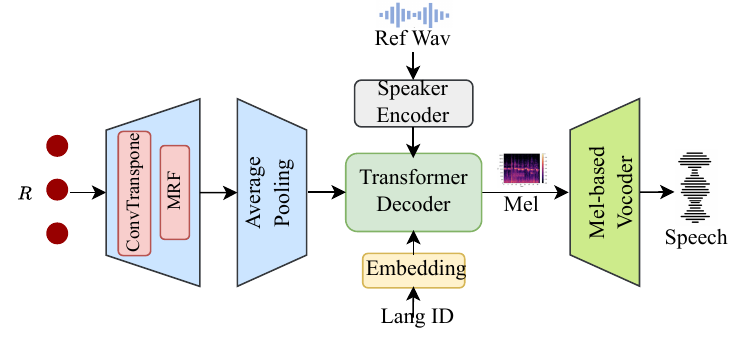}
  \caption{vec2wav with independent Mel-based vocoder.}
  \label{fig:3}
\end{figure}
\begin{figure}[t]
  \centering \includegraphics[width=\linewidth]{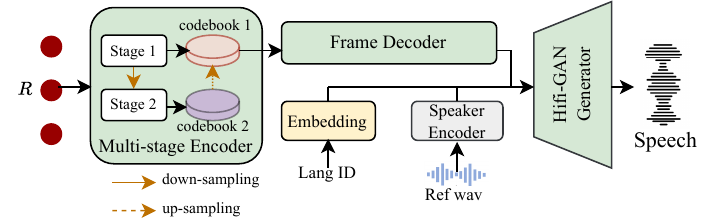}
  \caption{vec2wav without independent Mel-based vocoder.}
  \label{fig:4}
  \vspace{-0.4cm}
\end{figure}
In the process of converting discrete features $R$ into audio, we can still use a Mel-based vocoder, which requires us to first convert the discrete features $R$ into Mel spectrograms.

This vec2wav model consists of four parts: up-sampling, down-sampling, decoder, and pre-trained Mel-based vocoder as shown in Figure \ref{fig:3}. To learn Mel spectrograms from $R$, the duration model is no longer necessary. Although the sequences $R$ and $Mel$ have different “sampling rates,” they can be treated as being “uniformly aligned.” For example, if we adopt a 12.5 ms frame-shift to extract Mel spectrograms, considering the framerate of XLSR is 20ms,
the length ratio of {$\textcolor{black}{R:Mel:Speech}$} would be $5:8:1600$. In this work, we adopt an up-sampling factor of 8 and then down-sample by a factor of 5 to map discrete representations to Mel ones. Transposed convolution is used for up-sampling, followed by a Multi-Receptive Field Fusion (MRF) module, similar to the HiFi-GAN generator. With an up-sampling factor of 8, the configuration of the generator was changed for up-sampling rates to a sequence of (4, 2) with corresponding kernel sizes (12,8), while the hyper-parameters of residual blocks are the same as those in HiFi-GAN V1. For the down-sampling, we use average pooling. The decoder part is the same as that of txt2vec, which is an FFT. The decoder takes the output of average pooling and language and speaker embeddings to produce the Mel spectrograms. 
The optimization objective of this vec2mel is defined as:\textcolor{black}{
${\mathcal{L}_{Mel} = {\left\| \hat{Mel}-Mel \right\|}_2^2} $.}
\subsubsection{vec2wav model without Mel-based vocoder}
Instead of using an additional vocoder to convert Mel to a waveform, we also proposed a vec2wav model as in \cite{DBLP:conf/interspeech/SiuzdakDRJ22}, which directly maps discrete representations to a waveform. 
To learn the cross-lingual information from SSL representations at different time resolutions, we include a multi-stage multi-head codebook for the waveform modeling as \cite{guo2022multi,guo2022towards}.

The details of the architecture and operation of the multi-stage encoder are provided in \cite{guo2022multi,guo2022towards}. Here, we illustrate the audio waveform generation process with a 2-stage encoder as an example as shown in Figure \ref{fig:4}, which consists of the following steps: 1) The first-stage encoder receives discrete sequences $R$ and produces the first hidden representation sequences. These sequences are then downsampled and passed to the second-stage encoder to obtain the second hidden representation, which has a lower time resolution. 
2) The second hidden representation sequences are quantized using codebook 2 to obtain $z^{(2)}$. These sequences are then up-sampled and combined with the first hidden representation sequences to obtain $z^{(1)}$ using codebook 1. 
3) The residual output of $z^{(1)}$ and $z^{(2)}$ is used to reconstruct the waveform using the frame decoder and HiFi-GAN-based waveform generator.

To up-sample the audio from $R$, the generator's configuration was changed to a sequence of (5, 4, 4, 2, 2) with corresponding kernel sizes of (11, 8, 8, 4, 4). 

During the training process, a UnivNet \cite{jang2021univnet} discriminator is utilized for adversarial training of this model. The final loss is composed of three main components: waveform level loss $\mathcal{L}_{w}$, calculated using the HiFi-GAN loss function to compare the real waveform with the reconstructed waveform; frame-level loss $\mathcal{L}_{d}$, calculated from the mean squared error between two discrete representations; and VQ loss and stabilization loss, which are similar to those described in \cite{guo2022multi}.

\section{Experiments}
\label{sec:exp_setup}
\subsection{Dataset}
We investigated six languages including English (en), French (fr), German (ge), Portuguese (pt), Spanish (sp), and Swedish (sw), building a multilingual and multispeaker dataset to train the model. 
The multilingual and multispeaker dataset contains samples from two multispeaker datasets, multilingual LibriSpeech (MLS) \cite{Pratap2020MLSAL} and GlobalPhone (GLB) \cite{6639248}, and three single-speaker datasets, CSS10 \cite{park2019css10}, LJSpeech (LJS) \cite{ljspeech17}, and NST Swedish Speech \cite{NST}.
The MLS dataset comprises eight languages and is derived from Librivox audiobooks.
GlobalPhone is a database of multilingual read speech, including corresponding transcriptions and pronunciation dictionaries in 20 languages.
CSS10 is a compilation of monolingual single-speaker speech datasets for ten distinct languages.
LJSpeech is a single-speaker English dataset used in many studies. 
NST is a monolingual Swedish database designed for speech synthesis, initially created by Nordic Language Technology.

To ensure the balance of language, speakers, and gender, we screened the aforementioned databases and combined them into our multilingual and multispeaker database as shown in Table \ref{tab:data}. In addition to the two multispeaker databases MLS and GLB, for which most of the speakers have about 100 audio samples, we also selected 100 audio samples for each language from three single-speaker databases CSS10, LJS, and NST. The reason that we also selected some data from a single-speaker database is to aim to make a fairer comparison with single-speaker and single-language models in the future, but this is not included in this work.

To eliminate the potential impact of varying sampling rates as a confounding variable, we resampled all audio to a consistent 16 kHz rate and applied amplitude normalization using sv56 \cite{sv56}.
We set the frame and hop sizes to 1,024 and 200 (12.5 ms), respectively, to extract the Mel spectrograms from 16-kHz raw speech.
\begin{table}[t]
  \setlength\tabcolsep{3pt}
  \caption{Details of the training corpus. $\#$Spk represents the total number of speakers per gender and language. $\#$Sent represents the total number of utterances per gender and language. ``MLS'', ``GLB'' and ``Other'' represent a number of speakers selected from each of the MLS dataset, the Global Phone dataset and three single-speaker datasets, respectively.}
  \label{tab:data}
  \centering
  \begin{tabular}{llllllll}
\toprule
           \textbf{Lang}& \textbf{Gender} &\textbf{$\#$Spk} &\textbf{Dur(h)} &\textbf{$\#$Sent} &\textbf{MLS}  &\textbf{GLB} &\textbf{Other} \\ \midrule
\multirow{2}{*}{en}&Female &46 &12.06 &3,521 &45 &\phantom{0}0 &1(LJ)    \\ 
&Male &45 &11.87 &3,420 &45 &\phantom{0}0 &0 \\ \hline
\multirow{2}{*}{fr}&Female &45 &12.29 &4,661 &\phantom{0}0 &45 &0    \\ 
&Male &46 &11.88 &4,775 &\phantom{0}0 &45 &1(CSS10) \\ \hline
\multirow{2}{*}{ge}&Female &46 &11.92 &3,892 &38 &\phantom{0}7 &1(CSS10)   \\ 
&Male &45 &12.47 &6,576 &\phantom{0}0 &45 &0 \\ \hline
\multirow{2}{*}{pt}&Female &46 &\phantom{0}9.15 &3,596 &\phantom{0}3 &43 &0   \\ 
&Male &45 &\phantom{0}9.76 &3,793 &\phantom{0}4 &41 &0 \\ \hline
\multirow{2}{*}{es}&Female &45 &10.40 &3,206 &24 &21 &0   \\ 
&Male &46 &10.06 &3,168 &21 &24 &1(CSS10) \\ \hline
\multirow{2}{*}{sw}&Female &45 &\phantom{0}9.57 &4,866 &\phantom{0}0 &45 &0   \\ 
&Male &46 &\phantom{0}9.32 &4,776 &\phantom{0}0 &45 &1(NST)
\\\bottomrule
\end{tabular}
\end{table}
\begin{table}[t]
  \setlength\tabcolsep{8.5pt}
  \caption{Details of the pre-trained model.}
  \centering
  \begin{tabular}{llll}
\toprule
          \textbf{Name} &\textbf{Modality} &\textbf{Lang} &\textbf{Training data}  \\ \midrule
XLSR-53    &Audio    &53       &56K hours    \\
ECAPA-TDNN &Audio    &$>5$     &2794 hours     \\
XPhoneBERT &Text    &94        &330M sentences \\
\bottomrule
\label{tab:pretrained}
\end{tabular}
\end{table}
\subsection{Pre-trained Models}
In addition to utilizing the aforementioned six languages' paired datasets to train our TTS system directly, to incorporate more multilingual knowledge into our system, we also employed two pre-trained multilingual models, XLSR-53\footnote{https://huggingface.co/facebook/wav2vec2-large-xlsr-53} and XPhoneBERT\footnote{https://github.com/VinAIResearch/XPhoneBERT}, trained on audio-only and text-only data respectively, as shown in Table \ref{tab:pretrained}. For the speaker encoder, we use the publicly available ECAPA-TDNN model\footnote{https://huggingface.co/speechbrain/spkrec-ecapa-voxceleb}, which was trained with the multilingual VoxCeleb 1\&2 dataset.
\begin{table}[t]
 \setlength\tabcolsep{3.3pt}
  \caption{Experimental Setup. txt2intfeat and intfeat2wav represent text to intermediate features and intermediate features to waveform, respectively. SCL is speaker consistency loss. ECAPA and H/ASP are two different speaker encoders. Mel and SSL respectively represent the use of Mel spectrograms or self-supervised learning representations as intermediate features.}
  \label{tab:systems}
  \centering
  \begin{tabular}{lllllll}
\toprule
         {\textbf{\scriptsize SystemID}}&
         {\textbf{\scriptsize txt2intfeat}}&
         {\textbf{\scriptsize intfeat2wav}} &
         {\textbf{\scriptsize Spk}} &
         {\textbf{\scriptsize Mel}} &
         {\textbf{\scriptsize SSL}} &
         {\textbf{\scriptsize SCL}}
         \\
         \hline \hline
         \texttt{FSM1} & \scriptsize FastSpeech & \scriptsize HiFi-GAN & \scriptsize ECAPA & $\surd$  & &  \\
         \texttt{FSM2} & \scriptsize FastSpeech & \scriptsize vec2wavVQ (Mel) & \scriptsize ECAPA & $\surd$ & & \\
         \hline \hline
         \texttt{ZMM-TTS1} & \scriptsize txt2vec & \scriptsize vec2mel+HiFi-GAN & \scriptsize ECAPA & $\surd$  & $\surd$ & \\
         \texttt{ZMM-TTS2} & \scriptsize txt2vec & \scriptsize vec2wavVQ (R) & \scriptsize ECAPA  & & $\surd$ &  \\
         \hline \hline
         \texttt{YourTTS} & \multicolumn{2}{c}{\scriptsize VITS  } & \scriptsize H/ASP & & & $\surd$ \\
         \texttt{YourTTSE} & \multicolumn{2}{c}{\scriptsize VITS  } & \scriptsize ECAPA & & &$\surd$ \\
         \texttt{YourTTSW} & \multicolumn{2}{c}{\scriptsize VITS  } & \scriptsize ECAPA & & & \\
\bottomrule   
\end{tabular}
\vspace{-0.4cm}
\end{table}
\subsection{Experimental Setup and Model Architecture}
\begin{table*}[t]
\centering
\setlength\tabcolsep{1.9pt}
\caption{MOS, DMOS, and SECS results for seen speakers among different systems. The gray color scale indicates the relative value, and darker color indicates a better result. MOS and DMOS are reported with 95\% confidence intervals.}
\label{tab:seen}

\begin{tabular}{c|c|cc|cc|cc|cc|ccc|c}
\toprule
\multirow{3}{*}{Metrics}& \multirow{3}{*}{Lang} & \multicolumn{4}{c|}{Characters} & \multicolumn{4}{c|}{Pre-trained phoneme representations} & \multicolumn{3}{c|}{Characters}& \multirow{3}{*}{GT} \\
\cline{3-13}
& & \multicolumn{2}{c|}{FSM-based} & \multicolumn{2}{c|}{ZMM-TTS} & \multicolumn{2}{c|}{FSM-based} & \multicolumn{2}{c|}{ZMM-TTS} & \multicolumn{3}{c|}{YourTTS-based} & \\
\cline{3-13}
 & & \scriptsize \texttt{FSM1c}& \scriptsize \texttt{FSM2c}& \scriptsize \texttt{ZMM-TTS1c}& \scriptsize \texttt{ZMM-TTS2c} & \scriptsize \texttt{FSM1x}& \scriptsize \texttt{FSM2x}& \scriptsize \texttt{ZMM-TTS1x} & \scriptsize \texttt{ZMM-TTS2x} & \scriptsize \texttt{YourTTS} & \scriptsize \texttt{YourTTSE} & \scriptsize \texttt{YourTTSW} &\\ \hline

\multirow{6}{*}{MOS}
&en & \cellcolor[rgb]{1.00, 1.00, 1.00} {1.08±0.07} & \cellcolor[rgb]{0.98, 0.98, 0.98} {1.45±0.22} & \cellcolor[rgb]{0.86, 0.86, 0.86} {2.68±0.28} & \cellcolor[rgb]{0.83, 0.83, 0.83} {2.98±0.23} & \cellcolor[rgb]{0.98, 0.98, 0.98} {1.35±0.13} & \cellcolor[rgb]{0.96, 0.96, 0.96} {1.63±0.20} & \cellcolor[rgb]{0.83, 0.83, 0.83} {2.92±0.25} & \cellcolor[rgb]{0.78, 0.78, 0.78} {3.35±0.30} & \cellcolor[rgb]{0.92, 0.92, 0.92} {2.17±0.22} &-&-&\cellcolor[rgb]{0.67, 0.67, 0.67} {4.05±0.21}\\ 
&fr & \cellcolor[rgb]{0.97, 0.97, 0.97} {1.47±0.22} & \cellcolor[rgb]{0.93, 0.93, 0.93} {2.07±0.27} & \cellcolor[rgb]{0.84, 0.84, 0.84} {2.90±0.29} & \cellcolor[rgb]{0.64, 0.64, 0.64} {4.20±0.22} & \cellcolor[rgb]{0.96, 0.96, 0.96} {1.60±0.22} & \cellcolor[rgb]{0.88, 0.88, 0.88} {2.48±0.25} & \cellcolor[rgb]{0.81, 0.81, 0.81} {3.08±0.34} & \cellcolor[rgb]{0.62, 0.62, 0.62} {4.27±0.21} & \cellcolor[rgb]{0.75, 0.75, 0.75} {3.57±0.27} &-&-& \cellcolor[rgb]{0.59, 0.59, 0.59} {4.48±0.19}\\ 
&ge & \cellcolor[rgb]{0.96, 0.96, 0.96} {1.60±0.25} & \cellcolor[rgb]{0.93, 0.93, 0.93} {2.08±0.26} & \cellcolor[rgb]{0.82, 0.82, 0.82} {3.02±0.25} & \cellcolor[rgb]{0.69, 0.69, 0.69} {3.90±0.28} & \cellcolor[rgb]{0.96, 0.96, 0.96} {1.63±0.25} & \cellcolor[rgb]{0.92, 0.92, 0.92} {2.10±0.29} & \cellcolor[rgb]{0.86, 0.86, 0.86} {2.67±0.26} & \cellcolor[rgb]{0.73, 0.73, 0.73} {3.70±0.29} & \cellcolor[rgb]{0.83, 0.83, 0.83} {2.97±0.33} &-&-& \cellcolor[rgb]{0.69, 0.69, 0.69} {3.92±0.26}\\ 
&pt & \cellcolor[rgb]{0.98, 0.98, 0.98} {1.45±0.20} & \cellcolor[rgb]{0.96, 0.96, 0.96} {1.73±0.22} & \cellcolor[rgb]{0.85, 0.85, 0.85} {2.77±0.34} & \cellcolor[rgb]{0.81, 0.81, 0.81} {3.15±0.31} & \cellcolor[rgb]{0.97, 0.97, 0.97} {1.48±0.19} & \cellcolor[rgb]{0.93, 0.93, 0.93} {2.05±0.26} & \cellcolor[rgb]{0.87, 0.87, 0.87} {2.63±0.32} & \cellcolor[rgb]{0.81, 0.81, 0.81} {3.12±0.34} & \cellcolor[rgb]{0.89, 0.89, 0.89} {2.45±0.34} &-&-& \cellcolor[rgb]{0.61, 0.61, 0.61} {4.33±0.20}\\ 
&es & \cellcolor[rgb]{0.98, 0.98, 0.98} {1.37±0.14} & \cellcolor[rgb]{0.96, 0.96, 0.96} {1.68±0.21} & \cellcolor[rgb]{0.80, 0.80, 0.80} {3.18±0.26} & \cellcolor[rgb]{0.73, 0.73, 0.73} {3.70±0.21} & \cellcolor[rgb]{0.95, 0.95, 0.95} {1.83±0.17} & \cellcolor[rgb]{0.93, 0.93, 0.93} {2.02±0.22} & \cellcolor[rgb]{0.79, 0.79, 0.79} {3.25±0.26} & \cellcolor[rgb]{0.70, 0.70, 0.70} {3.87±0.24} & \cellcolor[rgb]{0.80, 0.80, 0.80} {3.17±0.29} &-&-& \cellcolor[rgb]{0.72, 0.72, 0.72} {3.75±0.28}\\ 
&sw & \cellcolor[rgb]{0.98, 0.98, 0.98} {1.33±0.16} & \cellcolor[rgb]{0.94, 0.94, 0.94} {1.88±0.21} & \cellcolor[rgb]{0.83, 0.83, 0.83} {2.98±0.29} & \cellcolor[rgb]{0.78, 0.78, 0.78} {3.32±0.34} & \cellcolor[rgb]{0.95, 0.95, 0.95} {1.77±0.24} & \cellcolor[rgb]{0.88, 0.88, 0.88} {2.47±0.25} & \cellcolor[rgb]{0.81, 0.81, 0.81} {3.08±0.28} & \cellcolor[rgb]{0.74, 0.74, 0.74} {3.65±0.32} & \cellcolor[rgb]{0.83, 0.83, 0.83} {2.93±0.30} &-&-& \cellcolor[rgb]{0.64, 0.64, 0.64} {4.18±0.25}\\
\hline
\multirow{6}{*}{DMOS}

&en & \cellcolor[rgb]{0.85, 0.85, 0.85} {3.80±0.28} & \cellcolor[rgb]{0.91, 0.91, 0.91} {3.40±0.32} & \cellcolor[rgb]{0.74, 0.74, 0.74} {4.35±0.27} & \cellcolor[rgb]{0.72, 0.72, 0.72} {4.42±0.29} & \cellcolor[rgb]{0.89, 0.89, 0.89} {3.55±0.30} & \cellcolor[rgb]{0.82, 0.82, 0.82} {3.95±0.24} & \cellcolor[rgb]{0.67, 0.67, 0.67} {4.60±0.20} & \cellcolor[rgb]{0.67, 0.67, 0.67} {4.63±0.23} & \cellcolor[rgb]{0.73, 0.73, 0.73} {4.37±0.26} &-&-& \cellcolor[rgb]{0.59, 0.59, 0.59} {4.92±0.09}\\ 
&fr & \cellcolor[rgb]{0.91, 0.91, 0.91} {3.40±0.39} & \cellcolor[rgb]{0.86, 0.86, 0.86} {3.70±0.39} & \cellcolor[rgb]{0.75, 0.75, 0.75} {4.28±0.34} & \cellcolor[rgb]{0.76, 0.76, 0.76} {4.27±0.34} & \cellcolor[rgb]{0.92, 0.92, 0.92} {3.33±0.38} & \cellcolor[rgb]{0.85, 0.85, 0.85} {3.77±0.42} & \cellcolor[rgb]{0.75, 0.75, 0.75} {4.32±0.33} & \cellcolor[rgb]{0.67, 0.67, 0.67} {4.62±0.24} & \cellcolor[rgb]{0.79, 0.79, 0.79} {4.08±0.34} &-&-& \cellcolor[rgb]{0.68, 0.68, 0.68} {4.58±0.24}\\ 
&ge & \cellcolor[rgb]{0.94, 0.94, 0.94} {3.20±0.37} & \cellcolor[rgb]{0.85, 0.85, 0.85} {3.80±0.34} & \cellcolor[rgb]{0.72, 0.72, 0.72} {4.42±0.24} & \cellcolor[rgb]{0.68, 0.68, 0.68} {4.58±0.19} & \cellcolor[rgb]{0.88, 0.88, 0.88} {3.58±0.37} & \cellcolor[rgb]{0.86, 0.86, 0.86} {3.73±0.38} & \cellcolor[rgb]{0.70, 0.70, 0.70} {4.50±0.24} & \cellcolor[rgb]{0.70, 0.70, 0.70} {4.52±0.25} & \cellcolor[rgb]{0.80, 0.80, 0.80} {4.05±0.31} &-&-& \cellcolor[rgb]{0.69, 0.69, 0.69} {4.55±0.24}\\ 
&pt & \cellcolor[rgb]{0.99, 0.99, 0.99} {2.73±0.36} & \cellcolor[rgb]{1.00, 1.00, 1.00} {2.63±0.39} & \cellcolor[rgb]{0.84, 0.84, 0.84} {3.83±0.33} & \cellcolor[rgb]{0.79, 0.79, 0.79} {4.10±0.30} & \cellcolor[rgb]{0.98, 0.98, 0.98} {2.83±0.35} & \cellcolor[rgb]{0.95, 0.95, 0.95} {3.12±0.37} & \cellcolor[rgb]{0.84, 0.84, 0.84} {3.82±0.35} & \cellcolor[rgb]{0.86, 0.86, 0.86} {3.73±0.37} & \cellcolor[rgb]{0.93, 0.93, 0.93} {3.30±0.39} &-&-& \cellcolor[rgb]{0.76, 0.76, 0.76} {4.23±0.35}\\ 
&es & \cellcolor[rgb]{0.87, 0.87, 0.87} {3.67±0.35} & \cellcolor[rgb]{0.87, 0.87, 0.87} {3.68±0.37} & \cellcolor[rgb]{0.75, 0.75, 0.75} {4.32±0.31} & \cellcolor[rgb]{0.74, 0.74, 0.74} {4.35±0.30} & \cellcolor[rgb]{0.77, 0.77, 0.77} {4.18±0.27} & \cellcolor[rgb]{0.81, 0.81, 0.81} {4.02±0.31} & \cellcolor[rgb]{0.72, 0.72, 0.72} {4.45±0.29} & \cellcolor[rgb]{0.71, 0.71, 0.71} {4.47±0.29} & \cellcolor[rgb]{0.65, 0.65, 0.65} {4.67±0.19} &-&-& \cellcolor[rgb]{0.62, 0.62, 0.62} {4.78±0.18}\\ 
&sw & \cellcolor[rgb]{0.95, 0.95, 0.95} {3.13±0.41} & \cellcolor[rgb]{0.92, 0.92, 0.92} {3.37±0.44} & \cellcolor[rgb]{0.82, 0.82, 0.82} {3.92±0.38} & \cellcolor[rgb]{0.75, 0.75, 0.75} {4.30±0.32} & \cellcolor[rgb]{0.94, 0.94, 0.94} {3.17±0.40} & \cellcolor[rgb]{0.88, 0.88, 0.88} {3.57±0.41} & \cellcolor[rgb]{0.84, 0.84, 0.84} {3.85±0.41} & \cellcolor[rgb]{0.73, 0.73, 0.73} {4.38±0.28} & \cellcolor[rgb]{0.82, 0.82, 0.82} {3.93±0.40} &-&-& \cellcolor[rgb]{0.75, 0.75, 0.75} {4.30±0.30}\\ \hline
\multirow{6}{*}{SECS}
&en & \cellcolor[rgb]{0.91, 0.91, 0.91} 0.792 & \cellcolor[rgb]{0.89, 0.89, 0.89} 0.813 & \cellcolor[rgb]{0.76, 0.76, 0.76} 0.907 & \cellcolor[rgb]{0.76, 0.76, 0.76} 0.906 & \cellcolor[rgb]{0.84, 0.84, 0.84} 0.852 & \cellcolor[rgb]{0.84, 0.84, 0.84} 0.855 & \cellcolor[rgb]{0.75, 0.75, 0.75} 0.912 & \cellcolor[rgb]{0.75, 0.75, 0.75} 0.914 & \cellcolor[rgb]{0.74, 0.74, 0.74} 0.922 & \cellcolor[rgb]{0.95, 0.95, 0.95} 0.749 & \cellcolor[rgb]{1.00, 1.00, 1.00} 0.688 & \cellcolor[rgb]{0.59, 0.59, 0.59} 0.999\\ 
&fr & \cellcolor[rgb]{0.73, 0.73, 0.73} 0.926 & \cellcolor[rgb]{0.74, 0.74, 0.74} 0.922 & \cellcolor[rgb]{0.68, 0.68, 0.68} 0.955 & \cellcolor[rgb]{0.68, 0.68, 0.68} 0.955 & \cellcolor[rgb]{0.72, 0.72, 0.72} 0.935 & \cellcolor[rgb]{0.74, 0.74, 0.74} 0.923 & \cellcolor[rgb]{0.68, 0.68, 0.68} 0.955 & \cellcolor[rgb]{0.67, 0.67, 0.67} 0.958 & \cellcolor[rgb]{0.66, 0.66, 0.66} 0.964 & \cellcolor[rgb]{0.86, 0.86, 0.86} 0.841 & \cellcolor[rgb]{0.88, 0.88, 0.88} 0.823 & \cellcolor[rgb]{0.59, 0.59, 0.59} 0.999\\ 
&ge & \cellcolor[rgb]{0.81, 0.81, 0.81} 0.874 & \cellcolor[rgb]{0.83, 0.83, 0.83} 0.861 & \cellcolor[rgb]{0.69, 0.69, 0.69} 0.949 & \cellcolor[rgb]{0.69, 0.69, 0.69} 0.949 & \cellcolor[rgb]{0.77, 0.77, 0.77} 0.901 & \cellcolor[rgb]{0.79, 0.79, 0.79} 0.889 & \cellcolor[rgb]{0.69, 0.69, 0.69} 0.947 & \cellcolor[rgb]{0.69, 0.69, 0.69} 0.950 & \cellcolor[rgb]{0.70, 0.70, 0.70} 0.943 & \cellcolor[rgb]{0.85, 0.85, 0.85} 0.847 & \cellcolor[rgb]{0.90, 0.90, 0.90} 0.807 & \cellcolor[rgb]{0.59, 0.59, 0.59} 0.999\\ 
&pt & \cellcolor[rgb]{0.85, 0.85, 0.85} 0.848 & \cellcolor[rgb]{0.86, 0.86, 0.86} 0.839 & \cellcolor[rgb]{0.71, 0.71, 0.71} 0.936 & \cellcolor[rgb]{0.71, 0.71, 0.71} 0.940 & \cellcolor[rgb]{0.78, 0.78, 0.78} 0.891 & \cellcolor[rgb]{0.82, 0.82, 0.82} 0.866 & \cellcolor[rgb]{0.71, 0.71, 0.71} 0.937 & \cellcolor[rgb]{0.69, 0.69, 0.69} 0.946 & \cellcolor[rgb]{0.71, 0.71, 0.71} 0.939 & \cellcolor[rgb]{0.90, 0.90, 0.90} 0.802 & \cellcolor[rgb]{0.94, 0.94, 0.94} 0.765 & \cellcolor[rgb]{0.59, 0.59, 0.59} 0.999\\ 
&es & \cellcolor[rgb]{0.77, 0.77, 0.77} 0.897 & \cellcolor[rgb]{0.80, 0.80, 0.80} 0.882 & \cellcolor[rgb]{0.71, 0.71, 0.71} 0.936 & \cellcolor[rgb]{0.70, 0.70, 0.70} 0.941 & \cellcolor[rgb]{0.75, 0.75, 0.75} 0.917 & \cellcolor[rgb]{0.77, 0.77, 0.77} 0.897 & \cellcolor[rgb]{0.72, 0.72, 0.72} 0.934 & \cellcolor[rgb]{0.70, 0.70, 0.70} 0.942 & \cellcolor[rgb]{0.69, 0.69, 0.69} 0.946 & \cellcolor[rgb]{0.87, 0.87, 0.87} 0.830 & \cellcolor[rgb]{0.91, 0.91, 0.91} 0.797 & \cellcolor[rgb]{0.59, 0.59, 0.59} 0.999\\ 
&sw & \cellcolor[rgb]{0.81, 0.81, 0.81} 0.876 & \cellcolor[rgb]{0.81, 0.81, 0.81} 0.877 & \cellcolor[rgb]{0.71, 0.71, 0.71} 0.940 & \cellcolor[rgb]{0.70, 0.70, 0.70} 0.943 & \cellcolor[rgb]{0.78, 0.78, 0.78} 0.895 & \cellcolor[rgb]{0.77, 0.77, 0.77} 0.901 & \cellcolor[rgb]{0.70, 0.70, 0.70} 0.941 & \cellcolor[rgb]{0.69, 0.69, 0.69} 0.947 & \cellcolor[rgb]{0.70, 0.70, 0.70} 0.944 & \cellcolor[rgb]{0.89, 0.89, 0.89} 0.810 & \cellcolor[rgb]{0.94, 0.94, 0.94} 0.772 & \cellcolor[rgb]{0.59, 0.59, 0.59} 0.999\\ 
 \bottomrule
\end{tabular}
\end{table*}
\begin{table*}[ht]
\centering
\setlength\tabcolsep{1.9pt}
\caption{MOS, DMOS, and SECS results for unseen speakers among different systems. The gray color scale indicates the relative value, and darker color indicates a better result. MOS and DMOS are reported with 95\% confidence intervals.}
\label{tab:unseen}
\begin{tabular}{c|c|cc|cc|cc|cc|ccc|c}
\toprule
\multirow{3}{*}{Metrics}& \multirow{3}{*}{Lang} & \multicolumn{4}{c|}{Characters} & \multicolumn{4}{c|}{Pre-trained phoneme representations} & \multicolumn{3}{c|}{Characters}& \multirow{3}{*}{GT} \\
\cline{3-13}
& & \multicolumn{2}{c|}{FSM-based} & \multicolumn{2}{c|}{ZMM-TTS} & \multicolumn{2}{c|}{FSM-based} & \multicolumn{2}{c|}{ZMM-TTS} & \multicolumn{3}{c|}{YourTTS-based} & \\
\cline{3-13}
 & & \scriptsize \texttt{FSM1c}& \scriptsize \texttt{FSM2c}& \scriptsize \texttt{ZMM-TTS1c}& \scriptsize \texttt{ZMM-TTS2c} & \scriptsize \texttt{FSM1x}& \scriptsize \texttt{FSM2x}& \scriptsize \texttt{ZMM-TTS1x} &\scriptsize \texttt{ZMM-TTS2x} & \scriptsize \texttt{YourTTS} &\scriptsize \texttt{YourTTSE} & \scriptsize \texttt{YourTTSW} &\\ \hline
\multirow{6}{*}{MOS}&
 en & \cellcolor[rgb]{1.00, 1.00, 1.00} {1.15±0.15} & \cellcolor[rgb]{0.98, 0.98, 0.98} {1.45±0.19} & \cellcolor[rgb]{0.84, 0.84, 0.84} {2.87±0.22} & \cellcolor[rgb]{0.83, 0.83, 0.83} {2.98±0.24} & \cellcolor[rgb]{0.98, 0.98, 0.98} {1.47±0.19} & \cellcolor[rgb]{0.96, 0.96, 0.96} {1.67±0.19} & \cellcolor[rgb]{0.88, 0.88, 0.88} {2.53±0.24} & \cellcolor[rgb]{0.76, 0.76, 0.76} {3.48±0.26} & \cellcolor[rgb]{0.88, 0.88, 0.88} {2.58±0.25} &-&-& \cellcolor[rgb]{0.68, 0.68, 0.68} {3.98±0.26}\\ 
&fr & \cellcolor[rgb]{0.99, 0.99, 0.99} {1.35±0.13} & \cellcolor[rgb]{0.94, 0.94, 0.94} {1.98±0.24} & \cellcolor[rgb]{0.79, 0.79, 0.79} {3.27±0.30} & \cellcolor[rgb]{0.61, 0.61, 0.61} {4.32±0.18} & \cellcolor[rgb]{0.96, 0.96, 0.96} {1.70±0.21} & \cellcolor[rgb]{0.88, 0.88, 0.88} {2.57±0.25} & \cellcolor[rgb]{0.84, 0.84, 0.84} {2.90±0.31} & \cellcolor[rgb]{0.60, 0.60, 0.60} {4.43±0.19} & \cellcolor[rgb]{0.79, 0.79, 0.79} {3.28±0.30} &-&-& \cellcolor[rgb]{0.61, 0.61, 0.61} {4.38±0.18}\\ 
&ge & \cellcolor[rgb]{0.97, 0.97, 0.97} {1.55±0.24} & \cellcolor[rgb]{0.94, 0.94, 0.94} {2.03±0.28} & \cellcolor[rgb]{0.80, 0.80, 0.80} {3.22±0.31} & \cellcolor[rgb]{0.75, 0.75, 0.75} {3.58±0.34} & \cellcolor[rgb]{0.96, 0.96, 0.96} {1.72±0.27} & \cellcolor[rgb]{0.91, 0.91, 0.91} {2.27±0.28} & \cellcolor[rgb]{0.81, 0.81, 0.81} {3.17±0.27} & \cellcolor[rgb]{0.75, 0.75, 0.75} {3.55±0.33} & \cellcolor[rgb]{0.82, 0.82, 0.82} {3.05±0.33} &-&-& \cellcolor[rgb]{0.59, 0.59, 0.59} {4.47±0.20}\\ 
&pt & \cellcolor[rgb]{0.97, 0.97, 0.97} {1.53±0.22} & \cellcolor[rgb]{0.94, 0.94, 0.94} {1.97±0.20} & \cellcolor[rgb]{0.79, 0.79, 0.79} {3.27±0.29} & \cellcolor[rgb]{0.76, 0.76, 0.76} {3.52±0.29} & \cellcolor[rgb]{0.95, 0.95, 0.95} {1.82±0.23} & \cellcolor[rgb]{0.90, 0.90, 0.90} {2.33±0.29} & \cellcolor[rgb]{0.78, 0.78, 0.78} {3.33±0.26} & \cellcolor[rgb]{0.68, 0.68, 0.68} {4.00±0.24} & \cellcolor[rgb]{0.81, 0.81, 0.81} {3.15±0.26} &-&-& \cellcolor[rgb]{0.66, 0.66, 0.66} {4.07±0.25}\\ 
&es & \cellcolor[rgb]{0.99, 0.99, 0.99} {1.32±0.17} & \cellcolor[rgb]{0.96, 0.96, 0.96} {1.75±0.20} & \cellcolor[rgb]{0.79, 0.79, 0.79} {3.28±0.23} & \cellcolor[rgb]{0.69, 0.69, 0.69} {3.95±0.25} & \cellcolor[rgb]{0.96, 0.96, 0.96} {1.68±0.17} & \cellcolor[rgb]{0.87, 0.87, 0.87} {2.65±0.26} & \cellcolor[rgb]{0.81, 0.81, 0.81} {3.12±0.26} & \cellcolor[rgb]{0.72, 0.72, 0.72} {3.77±0.28} & \cellcolor[rgb]{0.76, 0.76, 0.76} {3.50±0.24} &-&-& \cellcolor[rgb]{0.69, 0.69, 0.69} {3.90±0.28}\\ 
&sw & \cellcolor[rgb]{0.97, 0.97, 0.97} {1.52±0.19} & \cellcolor[rgb]{0.95, 0.95, 0.95} {1.88±0.23} & \cellcolor[rgb]{0.86, 0.86, 0.86} {2.78±0.26} & \cellcolor[rgb]{0.71, 0.71, 0.71} {3.83±0.29} & \cellcolor[rgb]{0.95, 0.95, 0.95} {1.87±0.23} & \cellcolor[rgb]{0.88, 0.88, 0.88} {2.60±0.27} & \cellcolor[rgb]{0.83, 0.83, 0.83} {2.95±0.30} & \cellcolor[rgb]{0.72, 0.72, 0.72} {3.72±0.27} & \cellcolor[rgb]{0.91, 0.91, 0.91} {2.25±0.30} &-&- & \cellcolor[rgb]{0.76, 0.76, 0.76} {3.47±0.32}\\\hline
 \multirow{6}{*}{DMOS}
 &en & \cellcolor[rgb]{0.93, 0.93, 0.93} {2.87±0.27} & \cellcolor[rgb]{0.95, 0.95, 0.95} {2.73±0.33} & \cellcolor[rgb]{0.91, 0.91, 0.91} {3.07±0.40} & \cellcolor[rgb]{0.92, 0.92, 0.92} {2.93±0.39} & \cellcolor[rgb]{0.97, 0.97, 0.97} {2.52±0.30} & \cellcolor[rgb]{1.00, 1.00, 1.00} {2.13±0.32} & \cellcolor[rgb]{0.96, 0.96, 0.96} {2.60±0.40} & \cellcolor[rgb]{0.95, 0.95, 0.95} {2.70±0.40} & \cellcolor[rgb]{0.79, 0.79, 0.79} {3.85±0.36} &-&-& \cellcolor[rgb]{0.59, 0.59, 0.59} {4.82±0.12}\\ 
&fr & \cellcolor[rgb]{0.94, 0.94, 0.94} {2.85±0.40} & \cellcolor[rgb]{0.90, 0.90, 0.90} {3.15±0.39} & \cellcolor[rgb]{0.89, 0.89, 0.89} {3.20±0.42} & \cellcolor[rgb]{0.86, 0.86, 0.86} {3.40±0.41} & \cellcolor[rgb]{0.94, 0.94, 0.94} {2.77±0.38} & \cellcolor[rgb]{0.92, 0.92, 0.92} {2.97±0.43} & \cellcolor[rgb]{0.87, 0.87, 0.87} {3.35±0.44} & \cellcolor[rgb]{0.85, 0.85, 0.85} {3.48±0.39} & \cellcolor[rgb]{0.79, 0.79, 0.79} {3.87±0.38} &-&-& \cellcolor[rgb]{0.65, 0.65, 0.65} {4.55±0.24}\\ 
&ge & \cellcolor[rgb]{0.99, 0.99, 0.99} {2.30±0.35} & \cellcolor[rgb]{0.94, 0.94, 0.94} {2.83±0.43} & \cellcolor[rgb]{0.94, 0.94, 0.94} {2.82±0.40} & \cellcolor[rgb]{0.90, 0.90, 0.90} {3.15±0.41} & \cellcolor[rgb]{0.96, 0.96, 0.96} {2.58±0.39} & \cellcolor[rgb]{0.93, 0.93, 0.93} {2.87±0.40} & \cellcolor[rgb]{0.92, 0.92, 0.92} {2.95±0.41} & \cellcolor[rgb]{0.89, 0.89, 0.89} {3.17±0.42} & \cellcolor[rgb]{0.87, 0.87, 0.87} {3.35±0.39} &-&-& \cellcolor[rgb]{0.68, 0.68, 0.68} {4.40±0.31}\\ 
&pt & \cellcolor[rgb]{0.97, 0.97, 0.97} {2.48±0.35} & \cellcolor[rgb]{0.95, 0.95, 0.95} {2.75±0.39} & \cellcolor[rgb]{0.83, 0.83, 0.83} {3.63±0.43} & \cellcolor[rgb]{0.80, 0.80, 0.80} {3.78±0.38} & \cellcolor[rgb]{0.91, 0.91, 0.91} {3.02±0.37} & \cellcolor[rgb]{0.91, 0.91, 0.91} {3.03±0.37} & \cellcolor[rgb]{0.88, 0.88, 0.88} {3.25±0.40} & \cellcolor[rgb]{0.84, 0.84, 0.84} {3.55±0.39} & \cellcolor[rgb]{0.84, 0.84, 0.84} {3.57±0.38} &-&-& \cellcolor[rgb]{0.60, 0.60, 0.60} {4.77±0.19}\\ 
&es & \cellcolor[rgb]{0.89, 0.89, 0.89} {3.18±0.36} & \cellcolor[rgb]{0.86, 0.86, 0.86} {3.45±0.34} & \cellcolor[rgb]{0.75, 0.75, 0.75} {4.12±0.33} & \cellcolor[rgb]{0.77, 0.77, 0.77} {3.98±0.35} & \cellcolor[rgb]{0.83, 0.83, 0.83} {3.60±0.36} & \cellcolor[rgb]{0.84, 0.84, 0.84} {3.53±0.39} & \cellcolor[rgb]{0.78, 0.78, 0.78} {3.93±0.35} & \cellcolor[rgb]{0.75, 0.75, 0.75} {4.07±0.36} & \cellcolor[rgb]{0.72, 0.72, 0.72} {4.25±0.34} &-&-& \cellcolor[rgb]{0.62, 0.62, 0.62} {4.68±0.23}\\ 
&sw & \cellcolor[rgb]{0.94, 0.94, 0.94} {2.77±0.42} & \cellcolor[rgb]{0.93, 0.93, 0.93} {2.90±0.42} & \cellcolor[rgb]{0.86, 0.86, 0.86} {3.43±0.44} & \cellcolor[rgb]{0.83, 0.83, 0.83} {3.62±0.44} & \cellcolor[rgb]{0.93, 0.93, 0.93} {2.90±0.40} & \cellcolor[rgb]{0.91, 0.91, 0.91} {3.07±0.48} & \cellcolor[rgb]{0.86, 0.86, 0.86} {3.43±0.44} & \cellcolor[rgb]{0.78, 0.78, 0.78} {3.88±0.41} & \cellcolor[rgb]{0.84, 0.84, 0.84} {3.52±0.43} &-&-& \cellcolor[rgb]{0.60, 0.60, 0.60} {4.77±0.18}\\ 
  \hline
 \multirow{6}{*}{SECS}
&en & \cellcolor[rgb]{0.96, 0.96, 0.96} 0.681 & \cellcolor[rgb]{0.95, 0.95, 0.95} 0.707 & \cellcolor[rgb]{0.87, 0.87, 0.87} 0.791 & \cellcolor[rgb]{0.87, 0.87, 0.87} 0.789 & \cellcolor[rgb]{0.92, 0.92, 0.92} 0.739 & \cellcolor[rgb]{0.93, 0.93, 0.93} 0.729 & \cellcolor[rgb]{0.87, 0.87, 0.87} 0.788 & \cellcolor[rgb]{0.88, 0.88, 0.88} 0.783 & \cellcolor[rgb]{0.80, 0.80, 0.80} 0.857 & \cellcolor[rgb]{0.98, 0.98, 0.98} 0.656 & \cellcolor[rgb]{1.00, 1.00, 1.00} 0.618 & \cellcolor[rgb]{0.59, 0.59, 0.59} 0.999\\ 
&fr & \cellcolor[rgb]{0.78, 0.78, 0.78} 0.867 & \cellcolor[rgb]{0.79, 0.79, 0.79} 0.865 & \cellcolor[rgb]{0.73, 0.73, 0.73} 0.911 & \cellcolor[rgb]{0.74, 0.74, 0.74} 0.905 & \cellcolor[rgb]{0.77, 0.77, 0.77} 0.877 & \cellcolor[rgb]{0.75, 0.75, 0.75} 0.892 & \cellcolor[rgb]{0.72, 0.72, 0.72} 0.918 & \cellcolor[rgb]{0.73, 0.73, 0.73} 0.911 & \cellcolor[rgb]{0.69, 0.69, 0.69} 0.936 & \cellcolor[rgb]{0.85, 0.85, 0.85} 0.812 & \cellcolor[rgb]{0.87, 0.87, 0.87} 0.786 & \cellcolor[rgb]{0.59, 0.59, 0.59} 0.999\\ 
&ge & \cellcolor[rgb]{0.85, 0.85, 0.85} 0.811 & \cellcolor[rgb]{0.84, 0.84, 0.84} 0.819 & \cellcolor[rgb]{0.78, 0.78, 0.78} 0.873 & \cellcolor[rgb]{0.76, 0.76, 0.76} 0.890 & \cellcolor[rgb]{0.81, 0.81, 0.81} 0.842 & \cellcolor[rgb]{0.81, 0.81, 0.81} 0.843 & \cellcolor[rgb]{0.78, 0.78, 0.78} 0.873 & \cellcolor[rgb]{0.75, 0.75, 0.75} 0.893 & \cellcolor[rgb]{0.72, 0.72, 0.72} 0.913 & \cellcolor[rgb]{0.86, 0.86, 0.86} 0.796 & \cellcolor[rgb]{0.91, 0.91, 0.91} 0.747 & \cellcolor[rgb]{0.59, 0.59, 0.59} 0.999\\ 
&pt & \cellcolor[rgb]{0.84, 0.84, 0.84} 0.821 & \cellcolor[rgb]{0.85, 0.85, 0.85} 0.806 & \cellcolor[rgb]{0.76, 0.76, 0.76} 0.890 & \cellcolor[rgb]{0.75, 0.75, 0.75} 0.899 & \cellcolor[rgb]{0.82, 0.82, 0.82} 0.840 & \cellcolor[rgb]{0.83, 0.83, 0.83} 0.825 & \cellcolor[rgb]{0.76, 0.76, 0.76} 0.885 & \cellcolor[rgb]{0.76, 0.76, 0.76} 0.883 & \cellcolor[rgb]{0.73, 0.73, 0.73} 0.911 & \cellcolor[rgb]{0.94, 0.94, 0.94} 0.714 & \cellcolor[rgb]{0.96, 0.96, 0.96} 0.680 & \cellcolor[rgb]{0.59, 0.59, 0.59} 0.999\\ 
&es & \cellcolor[rgb]{0.78, 0.78, 0.78} 0.870 & \cellcolor[rgb]{0.79, 0.79, 0.79} 0.861 & \cellcolor[rgb]{0.74, 0.74, 0.74} 0.904 & \cellcolor[rgb]{0.72, 0.72, 0.72} 0.915 & \cellcolor[rgb]{0.77, 0.77, 0.77} 0.880 & \cellcolor[rgb]{0.76, 0.76, 0.76} 0.886 & \cellcolor[rgb]{0.74, 0.74, 0.74} 0.901 & \cellcolor[rgb]{0.72, 0.72, 0.72} 0.916 & \cellcolor[rgb]{0.70, 0.70, 0.70} 0.929 & \cellcolor[rgb]{0.89, 0.89, 0.89} 0.772 & \cellcolor[rgb]{0.92, 0.92, 0.92} 0.741 & \cellcolor[rgb]{0.59, 0.59, 0.59} 0.999\\ 
&sw & \cellcolor[rgb]{0.82, 0.82, 0.82} 0.837 & \cellcolor[rgb]{0.82, 0.82, 0.82} 0.836 & \cellcolor[rgb]{0.75, 0.75, 0.75} 0.896 & \cellcolor[rgb]{0.74, 0.74, 0.74} 0.906 & \cellcolor[rgb]{0.80, 0.80, 0.80} 0.858 & \cellcolor[rgb]{0.80, 0.80, 0.80} 0.857 & \cellcolor[rgb]{0.75, 0.75, 0.75} 0.900 & \cellcolor[rgb]{0.73, 0.73, 0.73} 0.911 & \cellcolor[rgb]{0.72, 0.72, 0.72} 0.917 & \cellcolor[rgb]{0.91, 0.91, 0.91} 0.747 & \cellcolor[rgb]{0.94, 0.94, 0.94} 0.711 & \cellcolor[rgb]{0.59, 0.59, 0.59} 0.999\\ 
 \bottomrule
\end{tabular}
\vspace{-0.4cm}
\end{table*}
The proposed ZMM-TTS model can be implemented in two ways: \texttt{ZMM-TTS1} and \texttt{ZMM-TTS2}, in accordance with the difference of the vec2wav model, which respectively represents the use of an independent Mel-based vocoder and without an independent vocoder.
We consider the FastSpeech-based model (\texttt{FSM}) for comparison. 
It also uses two methods for implementation: one is to use the HiFi-GAN vocoder, and the other is to use a vec2wavVQ (R)-like structure as the vocoder. In addition, we also compare with a zero-shot multilingual model, YourTTS, which is based on a fully end-to-end model, VITS.

In summary, the differences of the various TTS systems are listed in Table \ref{tab:systems}. Furthermore, for \texttt{ZMM-TTS} and \texttt{FSM}, we also compare different input representations, using different suffixes to indicate them. 
For example, \texttt{ZMM-TTS1c}, \texttt{ZMM-TTS1i} and \texttt{ZMM-TTS1x} indicate that the inputs to the \texttt{ZMM-TTS1} model are characters, IPA, and pre-trained phoneme representations, respectively. 
The implementation details are summarized as follows and each module is trained independently:
\begin{itemize}
\item{\textbf{HiFi-GAN:}}
As for waveform generation in \texttt{FSM1} and \texttt{ZMM-TTS1}, HiFi-GAN\footnote{https://github.com/jik876/hifi-gan} is chosen as our neural vocoder. This HiFi-GAN is trained for 2.5M steps on a single NVIDIA A100 GPU with a batch size of 16 using our six languages training data.
\item{\textbf{FastSpeech:}}
Similar to txt2vec in Section~\ref{sec:method}-A, we also train a baseline system like FastSpeech as an acoustic model. This FastSpeech has the same encoder, decoder speaker, and language representations as txt2vec, but the output is Mel representations rather than SSL ones. FastSpeech, txt2vec, and vec2mel were all trained for 1.2M steps on a single NVIDIA A100 GPU with a batch size of 16.
\textcolor{black}{Both FastSpeech and txt2vec were built from an open-source repository\footnote{https://github.com/keonlee9420/Comprehensive-Transformer-TTS}.}

As mentioned in Section~\ref{sec:method}-A-1,
when incorporating pre-trained phoneme representations, models utilize XPhoneBERT as the phoneme encoder. XPhoneBERT is kept fixed during the initial 25\% of the training steps and is subsequently updated during the remaining training steps.
\item{\textbf{vec2wavVQ (Mel):} 
For a fairer comparison, we also train a vocoder with a VQ codebook as in vec2wavVQ (R) in III.C. 
The vec2wavVQ (Mel) has the same multi-stage encoder, and frame decoder as vec2wavVQ (R). They both use 2-stage 4-head codebooks. 
The input of vec2wavVQ (Mel) is Mel representations rather than discrete ones as in vec2wavVQ (R). Vec2wavVQ (Mel) and vec2wavVQ (R) were all trained for 1M steps on a single NVIDIA A100 GPU with a batch size of 16.
}
\item{\textbf{YourTTS:}}
\texttt{YourTTS} is a multilingual system based on VITS that has achieved state-of-the-art (SOTA) results in zero-shot multispeaker TTS.
The original YourTTS implementation uses the H/ASP model \cite{heo2020clova} as a speaker encoder. 
Therefore, we also train a \texttt{YourTTSE}, which means \texttt{YourTTS} using the same speaker encoder ECAPA as our ZMM-TTS model. 
Furthermore, there is a speaker consistency loss (SCL) in \texttt{YourTTS} to improve speaker similarity. 
Therefore, we also train a \texttt{YourTTSW}, in which ``W'' means YourTTS without SCL, as in our proposed model. In \texttt{YourTTSW}, ECAPA is also used as a speaker encoder rather than H/ASP.
All these YourTTS-based models were trained for one million steps on a single NVIDIA A100 GPU with a batch size of 32, and we always select the best checkpoint based on the development set.
\end{itemize}
\subsection{Evaluation Methods}
\label{sec:evaluation_metric}
\subsubsection{Subjective Evaluation Methods}
We synthesized sentences from test sets in six languages for each TTS system. For each language, we select 6 (3 female, 3 male) seen speakers and 6 (3 female, 3 male) unseen speakers for the test set, and two sentences for each speaker. 
Ten native listeners per language participated in our listening tests.
Listeners evaluated one test utterance at a time, initially assigning a rating on a Likert scale from 1 to 5 for the Mean Opinion Score (MOS) to assess naturalness. Subsequently, they provided ratings for speaker similarity concerning a reference utterance using a Differential MOS (DMOS) scale, ranging from 1 (indicating a different speaker) to 5 (indicating the same speaker).
Reference utterances were chosen randomly from the original speech of the target speaker. 
Due to the high cost of subjective evaluation, only the original \texttt{YourTTS} is used as the \texttt{YourTTS} related baseline to test MOS and DMOS. Our preliminary experimental results also demonstrated that YourTTS achieves better performance than \texttt{YourTTSE} and \texttt{YourTTSW}.
\subsubsection{Objective Evaluation Methods}
To assess the similarity between the synthesized voice and the original speaker, we determine the Speaker Encoder Cosine Similarity (SECS) by measuring the cosine similarity between the speaker embeddings of two audio samples extracted from the speaker encoder. The SECS score falls within the range of -1 to 1, with a higher value indicating better speaker similarity. In accordance with prior research \cite{casanova2021sc, DBLP:conf/icml/CasanovaWSJGP22}, we compute the SECS using the speaker encoder from the Resemblyzer package\footnote{https://github.com/resemble-ai/Resemblyzer}, facilitating comparisons with those studies.

We also objectively evaluate performance by measuring the intelligibility of speech content using an ASR algorithm. We synthesized 2,000 sentences (1,000 with seen and unseen speaker embeddings, respectively) for each language. Note that the language of the target speaker and text are always consistent. Although our model also has the ability of cross-lingual synthesis, considering that there is no clear standard for the evaluation of cross-language synthesis, this paper only focuses on the intra-lingual scenarios.

The evaluation texts are from the CMU ARCTIC sentences\footnote{http://festvox.org/cmu\_arctic/cmuarctic.data} for English, test data in MLS for French, German, Portuguese, and Spanish, and test data in NST for Swedish. All synthesized sentences were sent to the Whisper\footnote{https://github.com/openai/whisper} model for ASR. 
We computed the character error rate (CER) between the input text and the ASR-produced transcripts.

\section{Results and analysis}
\label{sec:exp_result_1}
The evaluation results for speech naturalness and speaker similarity are shown in Tables \ref{tab:seen} and \ref{tab:unseen} for seen and unseen target speakers, respectively.

\subsection{Subjective Evaluation Results}
\begin{figure}[t]
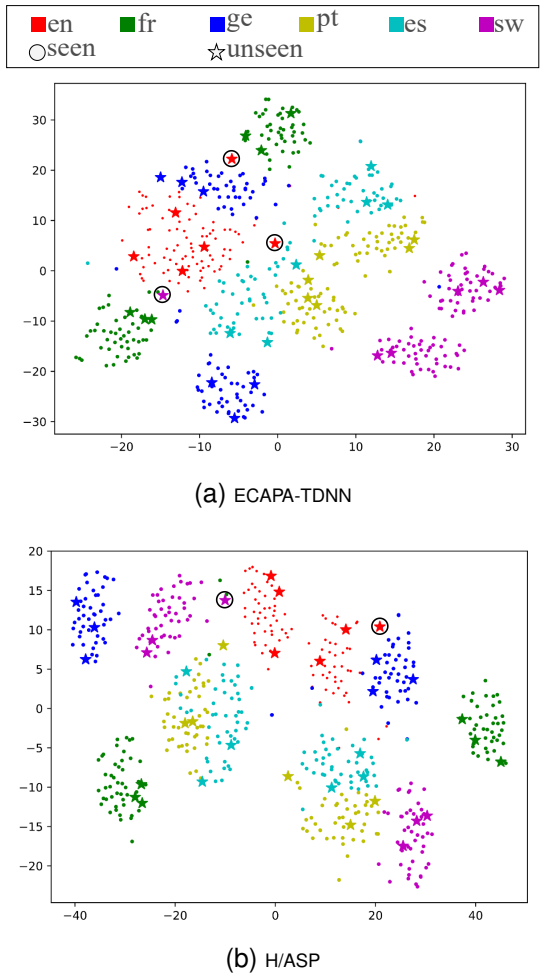

\centering
\subfloat[][\scriptsize ECAPA-TDNN]{\label{figure_ecapa_spk}\includegraphics[width=0.8\linewidth]{fig5_1212_a.pdf}}\hfill
\subfloat[][\scriptsize H/ASP]{\label{figure_yourtts_spk}\includegraphics[width=0.8\linewidth]{fig5_1214_b.pdf}}\hfill
\caption{Visualization of speaker embeddings extracted from ECAPA-TDNN and H/ASP.}
\label{fig:spk}
\end{figure}
\subsubsection{\textbf{Comparison between FSM and ZMM-TTS models}}
In terms of speech naturalness, our proposed systems ZMM-TTS1 and ZMM-TTS2 are significantly better, according to a Mann-Whitney U test given $\alpha= 0.05$ with Holm-Bonferroni correction, than the FastSpeech-based systems (FSM1 and FSM2) in each language on both seen and unseen speaker conditions.
Our training data come mainly from MLS and GLB. 
The sound quality of these datasets is significantly worse than several monolingual datasets such as LJSpeech \cite{ljspeech17} and the Chinese Standard Mandarin Speech Corpus \cite{biaobei} that are commonly used in speech synthesis. For example, most recordings of GLB were done in ordinary rooms rather than professional recording studios, and several audio samples contain some noise. 

As a currently popular method based on FastSpeech and HiFi-GAN, 
the baseline FastSpeech-based FSM model has achieved SOTA on many monolingual synthesis tasks while obtaining the worst results in speech naturalness in our multilingual datasets. This result indicates that for \texttt{FSM} models based on Mel spectrograms, the sound quality of synthesized audio may be limited due to the sound quality of the training data itself. 
On the other hand, since the intermediate features used by our ZMM-TTS model are extracted from SSL models trained on large-scale non-ideal data, the proposed system is considered to be less sensitive to the sound quality of the training data.
The reason that the proposed method \texttt{ZMTTS-2x} has higher MOS values than ground-truth audio on seen (Spanish) and unseen speakers (French and Swedish) would also be due to the fact that the sound quality of the ground-truth audio is not high.

In terms of speaker similarity, our proposed ZMM-TTS system achieves better DMOS than the FSM-based system. 
The speaker encoder used in the ZMM-TTS and FSM-based systems is the same. Also, the difference between the DMOS of the ZMM-TTS and FSM-based systems is smaller than their difference in MOS. 
This suggests that the improvement in naturalness using features extracted from SSL models is more prominent than the improvement in speaker similarity because discrete features may contain less speaker information.
\subsubsection{\textbf{Effectiveness of VQ in vec2wav}}
Another finding is that adding the VQ module in the process of converting the Mel spectrograms or discrete representations to a waveform can improve the naturalness of the audio.
The results in Tables \ref{tab:seen} and \ref{tab:unseen} show that \texttt{FSM2} outperformed \texttt{FSM1} on MOS. 
One potential reason is that the discrete representations in vec2wavVQ (Mel) may reduce some redundant information, such as noise, to some extent.
Similar findings are observed when comparing \texttt{ZMM-TTS1} and \texttt{ZMM-TTS2}. 
\texttt{ZMM-TTS2} has an additional VQ module and converts discrete representations directly to speech without an independent vocoder. Directly mapping discrete representations to a waveform instead of a Mel spectrogram prevents error propagation caused by inaccurate Mel spectrogram prediction in the vec2mel stage and hence, like the result of \texttt{FSM1} and \texttt{FSM2}, multi-stage discrete representations learned by additional VQ may also be helpful for improving sound quality.

In terms of speaker similarity, the use of VQ in vec2wav does not always result in an improvement. We also find that MOS is not always correlated with DMOS.
For example, as shown in the French results in Table \ref{tab:seen}, the naturalness of \texttt{ZMM-TTS2c} is significantly higher than that of \texttt{ZMM-TTS1c} (4.20 vs 2.90), while their DMOS values are very similar (4.27 vs 4.28). 
\subsubsection{\textbf{Comparison between YourTTS and ZMM-TTS models}}
On speech naturalness, the \texttt{YourTTS} method outperforms the two-stage \texttt{FSM} baseline while remaining worse than two of our proposed systems, \texttt{ZMM-TTS2c} and \texttt{ZMM-TTS2x}. We observed that the \texttt{YourTTS} model exhibits instability in its stochastic duration predictor, resulting in the production of unnatural durations for certain speakers and sentences.

On speaker similarity, \texttt{ZMM-TTS2c} and \texttt{ZMM-TTS2x} has better performance than \texttt{YourTTS} systems for the seen speaker condition in five languages (en, fr, ge, pt, and sw). However, under unseen conditions, \texttt{YourTTS} is better than our method in terms of DMOS in English, French, German and Spanish. This will be analyzed in detail in the next subsection.
\subsubsection{\textbf{Comparison between different input representations and languages}} 
Interestingly, determining which is better for speech naturalness, character-based inputs or pre-trained phoneme representation based inputs, depends on the specific language and synthesis systems. \texttt{FSM1x} and \texttt{FSM2x} always have a higher MOS than \texttt{FSM1c} and \texttt{FSM2c} among the FSM-based systems. 
Utilizing pre-trained phoneme representations as input and XPhoneBERT as the text encoder can enhance the quality of synthesized speech. However, for our proposed ZMM-TTS system, better naturalness is not always achieved with \texttt{ZMM-TTS1x} and \texttt{ZMM-TTS2x}. 
For example, in Table \ref{tab:unseen}, comparing the MOS of \texttt{ZMM-TTS2x} and \texttt{ZMM-TTS2c}, \texttt{ZMM-TTS2x} has achieved better results in English, French, and Portuguese, while \texttt{ZMM-TTS2c} has achieved better results in other languages. 
Furthermore, we observed that the MOS of \texttt{ZMM-TTS2x} is significantly better than \texttt{ZMM-TTS2c} in English. This may be because XPhoneBERT was pre-trained on a dataset with more English text than other languages.
We will investigate how the amount of pre-training text affects XPhoneBERT's representation performance in the future.

In terms of speaker similarity, different text representations have little impact. This result aligns with expectations as speaker and semantic information are not closely related.

\subsubsection{\textbf{Comparison between seen and unseen speakers}}
An interesting result is that the naturalness of the synthesized speech of unseen speakers is not worse than that of seen speakers.  
However, there was still a gap between seen and unseen speakers in terms of DMOS for most languages. Although the pre-training process for the speaker encoder involves more than 7,000 speakers, the TTS system is only trained on a few hundred speakers. 
It is still challenging to achieve the ideal zero-shot performance when training with limited multilingual data. 

\subsection{Objective Evaluation Results}

\begin{table*}[t]
\setlength\tabcolsep{1.1pt}
\caption{CER (\%) of ground-truth recordings (GT) and synthesized audio samples from various models.}
\label{tab:cer}
\begin{tabularx}{\textwidth}{c|cccc|cccc|cccc|ccc|c}
\toprule
\multirow{3}{*}{} &  \multicolumn{4}{c|}{Characters} &  \multicolumn{4}{c|}{IPA} & \multicolumn{4}{c|}{Pre-trained phoneme representations} & \multicolumn{3}{c|}{Characters}& \multirow{3}{*}{GT} \\

&  \multicolumn{2}{c|}{FSM-based} & \multicolumn{2}{c|}{ZMM-TTS} & \multicolumn{2}{c|}{FSM-based}  & \multicolumn{2}{c|}{ZMM-TTS} & \multicolumn{2}{c|}{FSM-based}  & \multicolumn{2}{c|}{ZMM-TTS} & \multicolumn{3}{c|}{YourTTS-based} & \\
 &  \scriptsize \texttt{FSM1c}& \scriptsize \texttt{FSM2c}& \scriptsize \texttt{ZMM-TTS1c}& \scriptsize \texttt{ZMM-TTS2c} & \scriptsize \texttt{FSM1i}& \scriptsize \texttt{FSM2i}&\scriptsize  \texttt{ZMM-TTS1i} &\scriptsize \texttt{ZMM-TTS2i} & \scriptsize \texttt{FSM1x}& \scriptsize \texttt{FSM2x}& \scriptsize \texttt{ZMM-TTS1x} &\scriptsize \texttt{ZMM-TTS2x} & \scriptsize \texttt{YourTTS} &\scriptsize  \texttt{YourTTSE} & \scriptsize \texttt{YourTTSW} &\\ \hline
  \multicolumn{4}{l}{\scriptsize \textit{Seen speakers}} \\
 
 \hline
 en& 9.30&11.30&8.70&\phantom{0}8.20&5.71&\phantom{0}6.98&\phantom{0}5.97&\phantom{0}5.66&2.90&3.70&5.10&5.20&\phantom{0}7.50&\phantom{0}7.22&\phantom{0}8.61& 0.44 \\

 fr& 4.43&\phantom{0}5.21&6.91&\phantom{0}6.72&6.07&\phantom{0}7.05&\phantom{0}8.49&\phantom{0}8.32&3.32&3.81&6.03&5.91&\phantom{0}6.92&\phantom{0}8.91&10.89&3.49\\

 ge& 3.82&\phantom{0}4.94&4.81&\phantom{0}4.62&5.26&\phantom{0}6.34&\phantom{0}5.63&\phantom{0}5.46&2.15&2.72&4.22&4.01&\phantom{0}7.21&\phantom{0}8.50&10.69 &2.84\\
  
pt& 2.62&\phantom{0}3.32&4.31&\phantom{0}4.17&3.63&\phantom{0}4.40&\phantom{0}5.26&\phantom{0}5.05&2.77&3.41&5.42&5.11&12.03&14.36&16.45&2.18 \\
es& 2.40&\phantom{0}3.23&3.41&\phantom{0}3.32&3.07&\phantom{0}3.71&\phantom{0}3.68&\phantom{0}3.59&1.58&2.01&2.67&2.73&\phantom{0}5.74&\phantom{0}7.78&\phantom{0}8.82&1.88 \\
 sw& 7.03&10.72&9.81&10.25&7.26&10.51&11.57&11.71&4.13&6.17&9.72&9.63&17.62&21.76&25.14
 & 2.64 \\

 \bottomrule

\multicolumn{4}{l}{\scriptsize \textit{Unseen speakers}} \\
 
 \hline
 
 en & 8.81&10.43&8.25&\phantom{0}7.62&5.14&\phantom{0}6.39&\phantom{0}5.59&\phantom{0}5.21&2.66&3.28&4.77&4.52&\phantom{0}7.23&\phantom{0}6.87&\phantom{0}8.61&0.44\\
 
 fr &4.42&\phantom{0}5.23&6.37&\phantom{0}6.02&5.88&\phantom{0}6.88&\phantom{0}7.74&\phantom{0}7.64&3.29&3.61&5.42&5.43&\phantom{0}7.02&\phantom{0}8.21&\phantom{0}9.46&3.49\\

 ge & 3.67&\phantom{0}4.42&4.01&\phantom{0}3.88&5.05&\phantom{0}5.86&\phantom{0}4.74&\phantom{0}4.69&2.17&2.49&3.52&3.44&\phantom{0}6.87&\phantom{0}8.36&\phantom{0}9.58&2.84\\

 pt & 2.56&\phantom{0}3.08&3.97&\phantom{0}3.63&3.55&\phantom{0}4.19&\phantom{0}4.56&\phantom{0}4.48&2.69&3.18&4.92&4.84&12.37&14.15&16.11&2.18\\

 es & 1.94&\phantom{0}2.57&2.83&\phantom{0}2.96&2.68&\phantom{0}3.17&\phantom{0}3.19&\phantom{0}3.22&1.56&1.88&2.44&2.32&\phantom{0}6.17&\phantom{0}7.55&\phantom{0}8.80&1.88\\

 sw & 6.61&10.42&8.53&\phantom{0}8.71&7.04&10.22&10.20&\phantom{0}9.78&3.79&5.42&8.61&8.82&17.92&21.18&24.75&2.64 \\ 
 \bottomrule
\end{tabularx}
\end{table*}

\subsubsection{\textbf{SECS analysis}}
First, we found very strong correlations ($r=0.807$), using the Pearson correlation coefficient (PCC), between the SECS and the DMOS values. 
Compared with the FSM-based model, ZMM-TTS performs better in both SECS and DMOS metrics for both seen and unseen speakers.
Compared with the \texttt{YourTTS} model, ZMM-TTS achieves similar SECS for seen speakers, but there are differences for unseen speakers, particularly English speakers.
One possible reason for this difference is that the original implementation of YourTTS utilizes SCL to enable generalization in the characteristics of unseen speakers. The different SECS results of \texttt{YourTTSE} and \texttt{YourTTSW} also demonstrate the impact of SCL.

To investigate whether the difference is related to language and the impact of two speaker encoders, we plotted speaker embeddings of seen and unseen speakers from natural speech via T-SNE\cite{van2008visualizing}. Figure \ref{fig:spk} presents the T-SNE visualization and shows the following:
Several unseen speakers are not close to the seen speakers in the training set. For example, there are two obvious outliers in the ECAPA-TDNN speaker embeddings of the English unseen speakers.
We conducted a study to calculate the SECS values of the six unseen English speakers whose speech was synthesized by the \texttt{ZMM-TTS1c} system. 
We found two English outlier points in Figure \ref{fig:spk}(a) with SECS values of 0.766 and 0.644, which were significantly lower than that for the speaker who was closer to the seen English speakers. 
This explains why the speaker similarity of our proposed \texttt{ZMM-TTS1c} for English unseen speakers is significantly lower than that of \texttt{YourTTS}.

Although there are many different types of SOTA neural speaker embeddings, currently, ZMM-TTS has only attempted using a popular speaker representation from \texttt{ECAPA-TDNN}, and we will explore the performance with different speaker embeddings such as H/ASP and SCL in future work.

\subsubsection{\textbf{CER analysis}}
Table \ref{tab:cer} summarizes the obtained CER. The first finding is that the \texttt{FSM1} and \texttt{FSM2} baseline models perform better, although their sound quality is worse than the other systems.
This demonstrates that the speech generated using Mel spectrograms as intermediate features is intelligible but not as natural sounding as when using SSL representations. Additionally, the speech synthesized through the Mel spectrograms tends to over-smooth the high-frequency region more than through SSL representations.

We also find that the CER is not always correlated with the MOS values. Although \texttt{YourTTS} CER is worse, the sound quality is obviously better compared with the FSM-based model. Furthermore, \texttt{YourTTS} CER in Portuguese and Swedish is significantly worse than in other languages. This also explains why the \texttt{YourTTS} system is significantly worse than our \texttt{ZMM-TTS} on MOS in these two languages. 
\texttt{YourTTS} only uses the character transcriptions rather than phonemes, which makes it more prone to mispronunciation issues as mentioned in \cite{DBLP:conf/icml/CasanovaWSJGP22}.

Additionally, we found that the pre-trained phoneme representation input system performance is the best, and IPA is only better than characters in English. A notable result is that compared with \texttt{FSM1}, the CER performance of \texttt{FSM2} has dropped significantly. 
For example, sometimes the pronunciation of the word ``advice'' in a sentence becomes incorrect and sounds more like the word ``addressed,'' resulting in ASR recognition errors.
This shows that learning a discrete representation from Mel spectrograms to speech conversion in vec2wavVQ (Mel) can improve the quality of speech naturalness, but some fine-grained information related to linguistic content may be lost.

\section{Low-resource scenarios}
\label{sec:exp_2}
\begin{table*}[t]
\centering
\setlength\tabcolsep{4pt}
\caption{Low-resource results. Total training data is 35 and 21 hours for Italian and Polish, respectively.}
\label{tab:low}

\begin{tabular}{c|c|c|c|c|ccc|c|c|c|ccc}
\toprule
&&\multicolumn{6}{c|}{Italian} &\multicolumn{6}{c}{Polish}\\
Size&Method&UTMOS&SECS&CER&\textcolor{black}{MOS}&\textcolor{black}{IMOS}&\textcolor{black}{DMOS}&UTMOS&SECS&CER&\textcolor{black}{MOS}&\textcolor{black}{IMOS}&\textcolor{black}{DMOS}\\
\hline
\multicolumn{10}{l}{Few-shot scenarios} \\
\hline
\multirow{4}{*}{\phantom{0}15m} \scriptsize &\texttt{FSM2x}&2.58 &0.85 &3.27 &2.51±0.14 &3.84±0.12 &2.62±0.27 &2.11 &0.89 &5.24 &2.03±0.17 &3.26±0.17 &2.29±0.28 \\
& \scriptsize \texttt{ZMM-TTS2x}&2.97 &0.91  &3.92 &3.33±0.17 &4.09±0.13 &2.95±0.28 &2.77 &0.93  &8.27 &3.54±0.23 &3.49±0.18 &2.91±0.30\\
&\scriptsize \texttt{FSM2c}&2.10 &0.78  &7.13 &1.62±0.14 &3.02±0.13 &2.29±0.25 &1.45 &0.83  &28.36 &1.41±0.16 &1.53±0.17 &1.81±0.23\\
&\scriptsize \texttt{ZMM-TTS2c}&2.86 &0.90 &6.07 &2.74±0.18 &3.25±0.14 &2.78±0.29 &2.55 &0.92  &13.34 &2.83±0.24 &2.34±0.17 &2.40±0.30\\
\hline
\multirow{2}{*}{\phantom{0}\phantom{0}5m}& \scriptsize \texttt{FSM2x}&2.60 &0.86  &3.39 
&2.73±0.16 &3.80±0.12 &2.90±0.27 &2.34 &0.90  &6.12 &2.52±0.21 &3.31±0.18 &2.46±0.30\\
&\scriptsize \texttt{ZMM-TTS2x}&2.92 &0.90  &4.02 &3.09±0.17 &3.88±0.13 &2.78±0.29 &2.69 &0.92  &10.01 &3.28±0.23 &3.28±0.20 &2.84±0.32\\
\hline
\multirow{2}{*}{2.5m}& \scriptsize \texttt{FSM2x}&2.48 &0.84  &3.12 &2.44±0.14 &3.62±0.12 &2.75±0.26 &2.22 &0.89  &11.29 &2.05±0.20 &2.56±0.18 &2.28±0.30\\
&\scriptsize \texttt{ZMM-TTS2x}&2.76 &0.89 &4.38 &2.90±0.17 &3.80±0.13 &2.67±0.29  &2.61 &0.91  & 14.64 &3.05±0.23 &2.83±0.17 &2.52±0.32\\
\hline
\multicolumn{10}{l}{Zero-shot scenarios} \\
\hline
\multirow{2}{*}{\phantom{0}\phantom{0}\phantom{0}0}&\scriptsize \texttt{FSM2x}&2.27 &0.73  &4.10 &1.66±0.15 &3.30±0.14 &1.43±0.15 &2.33 &0.77  &15.93 &1.60±0.18 &1.80±0.17 &1.34±0.16 \\
&\scriptsize \texttt{ZMM-TTS2x}&3.27 &0.83 &5.11 &2.88±0.20 &3.47±0.15 &1.52±0.25  &2.42 &0.85  &15.40 &2.95±0.26 &2.57±0.19 &2.05±0.29\\    
\hline
\multicolumn{10}{l}{High-resource baseline} \\
\hline
\phantom{0}\phantom{0}\phantom{0}1h&\scriptsize \texttt{YourTTS}&1.64 &0.74  &89.77 &-&-&-&1.40 &0.65  &88.86&-&-&-\\
35/21h&\scriptsize \texttt{YourTTS}&2.91 &0.96  &6.01 &-&-&- &1.70 &0.71  & 89.68&-&-&-\\
35/21h&\scriptsize \texttt{FSM2x}&2.27 &0.91  &3.29 &2.25±0.12 &4.21±0.14 &3.53±0.26 & 2.00 & 0.88 &2.89 &2.08±0.17 &3.79±0.21 &2.58±0.31\\
35/21h&\scriptsize \texttt{ZMM-TTS2x}&3.02 & 0.96  &3.76 &4.12±0.17 &4.32±0.20& 4.21±0.24 &2.87 &0.95 &4.49 &3.99±0.20 &4.28±0.20 &3.90±0.26\\
35/21h&\scriptsize \texttt{FSM2c}&2.17 &0.88&2.47 &2.01±0.15 &3.79±0.14 &2.99±0.27  & 1.79 & 0.85 &3.71 &1.78±0.17 &3.78±0.15 &2.49±0.31\\
35/21h&\scriptsize \texttt{ZMM-TTS2c}&2.92 & 0.96 &2.22 &3.77±0.17 
 &4.29±0.12 &4.19±0.24 & 2.70 &0.95  &2.42 &4.04±0.17 &4.29±0.14 &3.93±0.26\\
\hline
-&GT&2.94 &0.99  &3.17 &4.58±0.14 &4.80±0.08 &4.32±0.24 &2.82 &0.99 &3.95 &4.64±0.12 &4.88±0.08 &4.50±0.18\\
\bottomrule
\end{tabular}
\label{tab:result_low_resource}
\end{table*}
In addition to the high-resource languages, we also evaluate the performance of the proposed method in unseen low-resource language TTS scenarios to investigate language adaptability with limited training data.
\subsection{Experimental conditions}
\subsubsection{Dataset}
We chose Italian and Polish as two unseen languages for this experiment. In the language family tree \cite{do2022text}, Italian is closely related to French, Spanish, and Portuguese, all of which belong to the Romance family. Polish is relatively far away from our six high-resource languages and is from the Slavic language family.
We selected two speakers with ID \texttt{1595} and \texttt{6892} from MLS for our Italian and Polish data, respectively. 
\subsubsection{Implementation Details}
Considering the performance on six high-resource languages, we choose three models \texttt{FSM2}, \texttt{ZMM-TTS2} and \texttt{YourTTS} to implement on new languages. We have attempted to use both characters and pre-trained phonemes as input representations. 
\textcolor{black}{Low-resource language synthesis is investigated in two scenarios: few-shot and zero-shot. Few-shot means that the model is fine-tuned using the data from of the language before synthesing speech of that language.
In contrast, as mentioned in Section~\ref{sec:intro}, zero-shot means that our model performs inference on unseen languages without model fine-tuning.} 

{For the few-shot scenario}, we created training sets of different sizes (2.5, 5, and 15 mins of audio).
We use limited training data to finetune the model trained on six languages in Section~\ref{sec:exp_setup}.
For language embeddings, we reserve a free ID for new languages, which will be trained from scratch using fine-tuning data only.
Similarly, since different languages may use different characters, symbol embeddings for any characters unseen during pre-training must also be learned from fine-tuning data only. 
Specifically, in our experiments, we found that Italian and Polish had 3 and 14 unseen symbols, respectively, compared with the six pre-training languages.
All few-shot models are fine-tuned for 10,000 steps with a batch size of 16.
To compare the performance differences of the same language with high and low resources, we also perform fine-tuning with much larger datasets of 35h for Italian and 21h for Polish.

{For the zero-shot scenario}, to address the mismatch between language embeddings of seen and unseen languages, we removed the language embedding layer.
Therefore, the models tested on the zero-shot scenario are trained as in Section~\ref{sec:exp_setup} on six seen languages but without a language embedding layer.
Additionally, we used pre-trained phoneme representations as input, making it easily applicable to many unseen languages.
\subsubsection{Evaluation Methods}

\textbf{Objective Evaluation.}
We kept 100 sentences for the speech naturalness and speaker similarity tests and 1,000 sentences for ASR tests like those mentioned in Section~\ref{sec:evaluation_metric}. 
Furthermore, we adopted a publicly available automatic MOS (UTMOS) prediction model \cite{saeki2022utmos} to assess naturalness as in \cite{saeki2023learning,lee2023hierspeech++}.

\textcolor{black}{\textbf{Subjective Evaluation.} We synthesized eight sentences from test sets for each TTS system. 15 native listeners of each language participated in our listening tests. First, we employed the same evaluation as described in Section~\ref{sec:evaluation_metric} to assess MOS and DMOS. Furthermore, ensuring the intelligibility of generated speech is crucial, particularly in low-resource scenarios. To this aim, we also designed an intelligibility test (IMOS) in which listeners provided ratings for speech regarding the extent to which words were pronounced wrong or were not understandable. The scale of IMOS is also from  1 (indicating very bad) to 5 (indicating very good).}
\subsection{Experimental results and analysis}
The results on unseen languages are shown in Table~\ref{tab:result_low_resource}.
\subsubsection{Comparison between different input representations}
\textcolor{black}{In high-resource baselines, the intelligibility test results (CER/IMOS) of \texttt{ZMM-TTS2x} and \texttt{ZMM-TTS2c} were quite similar.}
However, when only 15 minutes of data were used for fine-tuning, the synthesized speech produced by \texttt{ZMM-TTS2x} was much more intelligible than that of \texttt{ZMM-TTS2c} in low-resource scenarios. The same phenomenon is also observed for \texttt{FSM2x} and \texttt{FSM2c}.
This result demonstrates that the pre-trained phoneme representations from XPhoneBert are more suitable for low-resource scenarios.
\subsubsection{Comparison of FSM and ZMM-TTS models}
\textcolor{black}{On both subjective and objective evaluation, the \texttt{ZMM-TTS2x} model produces significantly more natural speech (MOS/UTMOS) with better speaker similarity (DMOS/SECS) than the \texttt{FSM2x} model, both in high and low resource scenarios on two unseen languages.}
This result indicates that self-supervised features can better reconstruct sound quality and timbre compared with the Mel spectrograms, even with limited data. 
\textcolor{black}{Furthermore, on the basis of the subjective IMOS evaluations, in low-resource scenarios, the intelligibility of \texttt{ZMM-TTS2x} surpasses that of \texttt{FSM2x}. This demonstrates the advantage of our model in low-resource scenarios.}
\subsubsection{Comparison of ZMM-TTS and YourTTS models}
We can see that \texttt{YourTTS} cannot be fine-tuned with limited data from an unseen language. Even if one hour of data is used for fine-tuning, \texttt{YourTTS} still cannot synthesize understandable audio. 
Although promising results are evident in several language combinations with sufficient training data, \texttt{YourTTS} demonstrates limited efficacy when adapting to languages with limited available speakers. 

\subsubsection{Comparison of different data size}
As more training data is added, the synthesized speech quality improves, particularly in terms of speech intelligibility. 
\textcolor{black}{Furthermore, the results for speaker similarity (SECS/DMOS) of \texttt{ZMM-TTS2x} are similar across different quantities of training data (2.5, 5, and 15 mins).}
In general, with only 2.5 mins of speech from an unseen speaker in a new language, our fine-tuning of the \texttt{ZMM-TTS2x} still resulted in high speaker similarity in terms of SECS.
\subsubsection{Comparison of two languages}
We found that for Italian, using pre-trained phoneme representations as input, we were able to synthesize intelligible speech in zero-shot scenarios. In Polish, there is a significant gap between the zero-shot and few-shot scenarios, but both \texttt{FSM2x} and \texttt{ZMM-TTS2x} were able to generate intelligible speech utilizing a few minutes of data with pre-trained phoneme representations as input. 
The SSL speech-based features do not seem to provide an advantage over the Mel spectrograms, unlike in Italian. 
Although Italian and Polish are both included in the XLSR training data, the performance of SSL representations may be heavily affected by the similarity between the domains of the ZMM-TTS pre-training data and the fine-tuning data. Future work may include investigating the impact of language similarity on cross-lingual transfer.

\section{Conclusion}
\label{sec:conclusion}
In this paper, we propose a new method for generating multilingual speech using self-supervised discrete speech representations. We experimented with different input representations and incorporated a pre-trained multilingual phoneme encoder for multilingual tasks. Our experimental results show that our approach improves the speaker similarity and naturalness of synthetic speech in multilingual tasks, even for unseen speakers. Additionally, our framework achieves high intelligibility and speaker similarity in limited-data or zero-shot scenarios for a new language.

\textcolor{black}{As future work, we plan to apply this model to many more languages and explore advanced language and speaker adaptation strategies. To make this happen, we will also need to investigate flexible language representations to handle unseen languages more accurately than the language IDs used in the current system. }

\appendices
\section{\textcolor{black}{Comparison with large-scale TTS systems}}
\label{sec:appendix}
Considering the remarkable breakthroughs achieved by large-scale TTS models for monolingual speech synthesis, despite our model primarily targeting multilingual and low-resource scenarios, we still conducted comparative analyses from various perspectives on the English language, in contrast to these models.
\subsection{Experimental conditions}
\subsubsection{\textbf{Implementation details}} 
For a fair comparison, in addition to the ZMM-TTS2x model trained on six languages in Section~\ref{sec:exp_setup}, we also trained another model on the LibriTTS \cite{zen2019libritts} dataset.
We compared our ZMM-TTS2x with the following three strong zero-shot TTS baselines:
\begin{itemize} 
\item
\textbf{\texttt{VALL-E-X\cite{zhang2023speak}.}}
It is a multilingual version of VALL-E that also employs a combination of autoregressive and non-autoregressive methods for discrete token generation.
We use an open-source implementation and checkpoint \footnote{https://github.com/Plachtaa/VALL-E-X}. The training data for this model comprises three languages: Chinese (ch), English (en), and Japanese (jp).
\item \textbf{\texttt{HierSpeech++\cite{lee2023hierspeech++}.}}
This work is an extended version of HierSpeech \cite{lee2022hierspeech}. It uses a non-autoregressive model for continuous vector generation. 
There are three components in HierSpeech++: a hierarchical speech synthesizer, text-to-vec (TTV), and speech super-resolution (SpeechSR). 
We conducted experiments using two different pre-trained HierSpeech++ models, which were labeled as HierSpeech++(a) and HierSpeech++(b). The difference between them lies in the training data scale of the hierarchical speech synthesizer. HierSpeech++(a) used the complete LibriTTS train-clean data, while HierSpeech++(b) utilized LibriTTS train-clean-100 and 360 subsets. However, the training data for TTV in both HierSpeech++(a) and HierSpeech++(b) remained consistent, employing the complete LibriTTS train-clean data. Note that for purposes of fair comparison, we did not use the super-resolution model. We used the official code and checkpoint for the experiments \footnote{https://github.com/sh-lee-prml/HierSpeechpp}.

\item \textbf{\texttt{StyleTTS 2\cite{li2024styletts}.}}
It leverages style diffusion and adversarial training with large speech language models (12-layer WavLM pre-trained on 94k hours of data) to achieve human-level TTS synthesis.
We use the official code and checkpoint \footnote{https://github.com/yl4579/StyleTTS2}.
\end{itemize}

\subsubsection{\textbf{Evaluation dataset}}
We chose LibriSpeech \cite{panayotov2015librispeech} test-clean as our benchmark dataset for the zero-shot TTS task. This widely-used test set comprises 40 different speakers and 5.4 hours of speech. To conduct the benchmark experiments, we followed the approach described in \cite{ju2024naturalspeech}, and we randomly selected 25 sentences for each speaker from the LibriSpeech test-clean dataset.

\subsubsection{\textbf{Metrics}}
We evaluate the intelligibility of synthesized speech using WER as described in Section~\ref{sec:evaluation_metric}. 
Following previous works \cite{wang2023neural,ju2024naturalspeech}, we compute SECS using the SOTA speaker verification model, WavLM-Large\footnote{https://github.com/microsoft/UniSpeech/tree/main/\\downstreams/speaker\_verification}, to evaluate the speaker similarity,  enabling comparison with those studies.
We also use UTMOS like Section~\ref{sec:exp_2}.
In addition to evaluating the speech quality, we also measure the efficiency of the proposed model based on its Real-Time Factor (RTF), which is the time it takes to generate one second of audio on GPU and the number of parameters used in the model. We test the RTF on a single GPU (NVIDIA RTX 4090 with 24GB memory). 
Note that VALL-E-X and StyleTTS 2 generate audio using a sample rate of 24 kHz, while ZMM-TTS2x and HierSpeech++ use a sample rate of 16 kHz. To ensure consistency across all audio, we resampled it to 16 kHz and applied amplitude normalization using sv56 before conducting the WER, SECS, and UTMOS tests. 

\begin{table*}[tbp]
\centering
\caption{\textcolor{black}{Evaluation results for ZMM-TTS and recent large-scale TTS systems on LibriSpeech test-clean. LT-460 denotes LibriTTS train-clean-100 and 360 subsets, and LT-960 additionally utilizes a train-other-500 subset with LT-460. AL-1\&3 denotes AISHELL-1 and  AISHELL-3, and JP-CV denotes the Japanese Common Voice.}}
\setlength\tabcolsep{4pt}
\label{tab:large}
\begin{tabular}{l|ccrr|rrrrrr}
\toprule
\multirow{2}{*}{Method} & \multicolumn{4}{c|}{Training corpus} & \multicolumn{5}{c}{Evaluation metric} \\
\cline{2-10}
 &Datasets & Language& Speaker &Hour &WER &SECS &UTMOS &Params &RTF \\
\midrule
\texttt{VALL-E-X} & LT-960, AL-1\&3, JP-CV, others & en, jp, ch & $>3,122$ &$1,739$ & 26.77  &0.512 & 3.29 &395M &5.917\\
\texttt{HierSpeech++(a)} & LT-960 &en & 2,311 & 555 & \textbf{2.03} & 0.591& \textbf{4.40} &204M &0.217 \\
\texttt{HierSpeech++(b)} & LT-460 &en & 1,151 & 245 & 2.17 &0.555& 4.38 &204M &0.222 \\
\texttt{StyleTTS 2} &LT-460 &en &1,151& 245 &3.06 &0.455 &4.23 &191M &0.070 \\
\hline
\texttt{ZMM-TTS2x(a)} & MLS, GLB, LJ, CSS10, NST& en, fr, ge, pt, es, sw & 546 &130 &4.06 &0.432 &3.89 &\textbf{167M} &\textbf{0.003} \\
\texttt{ZMM-TTS2x(b)} & LT-960& en& 2,311 &555 & 2.37&\textbf{0.644}&4.07 &\textbf{167M} &\textbf{0.003} \\
\hline \hline
\texttt{Ground-truth} & -  &-   &-  &- & 2.14  &- &4.13 &- &- \\ 
\bottomrule
\end{tabular}
\end{table*}

\subsection{Experimental results and analysis}
Table~\ref{tab:large} contains information on the efficiency and quality of all models being compared. 
\subsubsection{\textbf{SECS analysis}}
We found that models \texttt{VALL-E-X}, \texttt{HierSpeech++(a)}, and \texttt{HierSpeech++(b)} outperform our model \texttt{ZMM-TTS2x (a)} in terms of unseen speaker similarity on LibriSpeech test sets. This gap may be due to differences in the scale of training data.
\texttt{ZMM-TTS2x (a)} training corpus includes only 546 speakers, while other baseline systems have over 1,000 or 2,000 speakers in their training data. When reducing the scale of training data, the SECS of \texttt{HierSpeech++(b)} has a significant decline compared with \texttt{HierSpeech++(a)}. 
With the same data scale, our proposed \texttt{ZMM-TTS2x (b)} achieved the best speaker similarity SECS compared with other recent zero-shot models.
This demonstrates that our proposed model, which utilizes a larger amount of data, scales in performance and can generalize to unseen speakers.

\subsubsection{\textbf{UTMOS analysis}}
After analyzing the scores predicted by the automatic MOS evaluation model, we noticed a correspondence between the quality of generated audio and the training corpus. Models trained on a clean TTS database LT-960 or LT-640 all achieved better UTMOS values. The training corpora for both the \texttt{ZMM-TTS2x (a)} model and \texttt{VALL-E-X} model do not exclusively consist of high-quality clean audio. They also contain some data that includes noticeable noise, such as the GlobalPhone in \texttt{ZMM-TTS2x (a)} and other databases in \texttt{VALL-E-X}.

\subsubsection{\textbf{CER analysis}}
We have observed that the speech intelligibility achieved by the \texttt{VALL-E-X} model is the worst among all the models. This is primarily due to the autoregressive process of the model, which often results in synthesized speech containing errors such as skipped words and repeated words. On the other hand, our \texttt{ZMM-TTS2x (a)} model, despite being trained on a smaller corpus and multiple languages, has achieved WER results in English that are comparable with other models. This indicates the robustness of our model.
\subsubsection{\textbf{Latency analysis}}
Undoubtedly, the autoregressive process of \texttt{VALL-E-X} results in the slowest synthesis speed, making real-time synthesis challenging to achieve on GPUs. Due to its non-autoregressive structure, \texttt{ZMM-TTS2x} is capable of real-time high-speed synthesis on GPU.
Although \texttt{HierSpeech++} is also non-autoregressive, the iterative process of the diffusion model still impacts the synthesis speed.
Additionally, \texttt{ZMM-TTS2x} has around 167M parameters, thus being smaller than other TTS systems we compare.

Note that the baseline models focus solely on resource-rich languages such as English. In contrast, our model has strong language adaptability, making it more suitable for low-resource languages. 

\bibliographystyle{IEEEtran}
\bibliography{mybib}

\begin{thebibliography}{10}
\providecommand{\url}[1]{#1}
\csname url@samestyle\endcsname
\providecommand{\newblock}{\relax}
\providecommand{\bibinfo}[2]{#2}
\providecommand{\BIBentrySTDinterwordspacing}{\spaceskip=0pt\relax}
\providecommand{\BIBentryALTinterwordstretchfactor}{4}
\providecommand{\BIBentryALTinterwordspacing}{\spaceskip=\fontdimen2\font plus
\BIBentryALTinterwordstretchfactor\fontdimen3\font minus
  \fontdimen4\font\relax}
\providecommand{\BIBforeignlanguage}[2]{{%
\expandafter\ifx\csname l@#1\endcsname\relax
\typeout{** WARNING: IEEEtran.bst: No hyphenation pattern has been}%
\typeout{** loaded for the language `#1'. Using the pattern for}%
\typeout{** the default language instead.}%
\else
\language=\csname l@#1\endcsname
\fi
#2}}
\providecommand{\BIBdecl}{\relax}
\BIBdecl

\bibitem{DBLP:conf/interspeech/WangSSWWJYXCBLA17}
Y.~Wang, R.~J. Skerry{-}Ryan, D.~Stanton, Y.~Wu, R.~J. Weiss, N.~Jaitly,
  Z.~Yang, Y.~Xiao, Z.~Chen, S.~Bengio, Q.~V. Le, Y.~Agiomyrgiannakis,
  R.~Clark, and R.~A. Saurous, ``Tacotron: {Towards} end-to-end speech
  synthesis,'' in \emph{Proc. Interspeech}.\hskip 1em plus 0.5em minus
  0.4em\relax {ISCA}, 2017, pp. 4006--4010.

\bibitem{DBLP:conf/iclr/0006H0QZZL21}
Y.~Ren, C.~Hu, X.~Tan, T.~Qin, S.~Zhao, Z.~Zhao, and T.~Liu, ``Fastspeech 2:
  {Fast} and high-quality end-to-end text to speech,'' in \emph{Proc. ICLR},
  2021.

\bibitem{DBLP:conf/icassp/ShenPWSJYCZWRSA18}
J.~Shen, R.~Pang, R.~J. Weiss, M.~Schuster, N.~Jaitly, Z.~Yang, Z.~Chen,
  Y.~Zhang, Y.~Wang, R.~Ryan, R.~A. Saurous, Y.~Agiomyrgiannakis, and Y.~Wu,
  ``Natural {TTS} synthesis by conditioning {WaveNet} on {Mel} spectrogram
  predictions,'' in \emph{Proc. ICASSP}.\hskip 1em plus 0.5em minus 0.4em\relax
  {IEEE}, 2018, pp. 4779--4783.

\bibitem{tan2022naturalspeech}
X.~Tan, J.~Chen, H.~Liu, J.~Cong, C.~Zhang, Y.~Liu, X.~Wang, Y.~Leng, Y.~Yi,
  L.~He \emph{et~al.}, ``{NaturalSpeech}: End-to-end text to speech synthesis
  with human-level quality,'' \emph{arXiv preprint arXiv:2205.04421}, 2022.

\bibitem{saeki2023learning}
T.~Saeki, S.~Maiti, X.~Li, S.~Watanabe, S.~Takamichi, and H.~Saruwatari,
  ``Learning to speak from text: {Zero}-shot multilingual text-to-speech with
  unsupervised text pretraining,'' in \emph{Proc. {{IJCAI}}}, 2023.

\bibitem{DBLP:conf/interspeech/ZhangWZWCSJRR19}
Y.~Zhang, R.~J. Weiss, H.~Zen, Y.~Wu, Z.~Chen, R.~J. Skerry{-}Ryan, Y.~Jia,
  A.~Rosenberg, and B.~Ramabhadran, ``Learning to speak fluently in a foreign
  language: Multilingual speech synthesis and cross-language voice cloning,''
  in \emph{Proc. Interspeech}.\hskip 1em plus 0.5em minus 0.4em\relax {ISCA},
  2019, pp. 2080--2084.

\bibitem{DBLP:conf/interspeech/NekvindaD20}
T.~Nekvinda and O.~Dusek, ``One model, many languages: Meta-learning for
  multilingual text-to-speech,'' in \emph{Proc. Interspeech}.\hskip 1em plus
  0.5em minus 0.4em\relax ISCA, 2020, pp. 2972--2976.

\bibitem{DBLP:conf/icml/CasanovaWSJGP22}
E.~Casanova, J.~Weber, C.~D. Shulby, A.~C. J{\'{u}}nior, E.~G{\"{o}}lge, and
  M.~A. Ponti, ``{YourTTS}: Towards zero-shot multi-speaker {TTS} and zero-shot
  voice conversion for everyone,'' in \emph{Proc. ICML}, vol. 162.\hskip 1em
  plus 0.5em minus 0.4em\relax {PMLR}, 2022, pp. 2709--2720.

\bibitem{jia2021png}
Y.~Jia, H.~Zen, J.~Shen, Y.~Zhang, and Y.~Wu, ``{PnG BERT}: Augmented {BERT} on
  phonemes and graphemes for neural {TTS},'' \emph{Proc. Interspeech}, pp.
  151--155, 2021.

\bibitem{zhang2022mixed}
G.~Zhang, K.~Song, X.~Tan, D.~Tan, Y.~Yan, Y.~Liu, G.~Wang, W.~Zhou, T.~Qin,
  T.~Lee \emph{et~al.}, ``Mixed-phoneme {BERT}: Improving bert with mixed
  phoneme and sup-phoneme representations for text to speech,'' in \emph{Proc.
  Interspeech}, 2022, pp. 456--460.

\bibitem{li2023phoneme}
Y.~A. Li, C.~Han, X.~Jiang, and N.~Mesgarani, ``Phoneme-level {BERT} for
  enhanced prosody of text-to-speech with grapheme predictions,'' in
  \emph{Proc. ICASSP}.\hskip 1em plus 0.5em minus 0.4em\relax IEEE, 2023, pp.
  1--5.

\bibitem{xphonebert}
L.~T. Nguyen, T.~Pham, and D.~Q. Nguyen, ``{XPhoneBERT}: A pre-trained
  multilingual model for phoneme representations for text-to-speech,'' in
  \emph{Proc. Interspeech}, 2023, pp. 5506--5510.

\bibitem{NEURIPS2018_6832a7b2}
Y.~Jia, Y.~Zhang, R.~Weiss, Q.~Wang, J.~Shen, F.~Ren, z.~Chen, P.~Nguyen,
  R.~Pang, I.~Lopez~Moreno, and Y.~Wu, ``Transfer learning from speaker
  verification to multispeaker text-to-speech synthesis,'' in \emph{Proc.
  NIPS}, vol.~31, 2018, pp. 4480--4490.

\bibitem{9054535}
E.~Cooper, C.-I. Lai, Y.~Yasuda, F.~Fang, X.~Wang, N.~Chen, and J.~Yamagishi,
  ``Zero-shot multi-speaker text-to-speech with state-of-the-art neural speaker
  embeddings,'' in \emph{Proc. ICASSP}, 2020, pp. 6184--6188.

\bibitem{DBLP:conf/nips/BaevskiZMA20}
A.~Baevski, Y.~Zhou, A.~Mohamed, and M.~Auli, ``Wav2vec 2.0: {A} framework for
  self-supervised learning of speech representations,'' in \emph{Proc. NIPS},
  2020, pp. 12\,449--12\,460.

\bibitem{hsu2021hubert}
W.-N. Hsu, B.~Bolte, Y.-H.~H. Tsai, K.~Lakhotia, R.~Salakhutdinov, and
  A.~Mohamed, ``{HuBERT}: Self-supervised speech representation learning by
  masked prediction of hidden units,'' \emph{IEEE/ACM Transactions on Audio,
  Speech, and Language Processing}, vol.~29, pp. 3451--3460, 2021.

\bibitem{9814838}
S.~Chen, C.~Wang, Z.~Chen, Y.~Wu, S.~Liu, Z.~Chen, J.~Li, N.~Kanda,
  T.~Yoshioka, X.~Xiao, J.~Wu, L.~Zhou, S.~Ren, Y.~Qian, Y.~Qian, J.~Wu,
  M.~Zeng, X.~Yu, and F.~Wei, ``{WavLM}: Large-scale self-supervised
  pre-training for full stack speech processing,'' \emph{IEEE Journal of
  Selected Topics in Signal Processing}, vol.~16, no.~6, pp. 1505--1518, 2022.

\bibitem{9746223}
B.~Thomas, S.~Kessler, and S.~Karout, ``Efficient adapter transfer of
  self-supervised speech models for automatic speech recognition,'' in
  \emph{Proc. ICASSP}, 2022, pp. 7102--7106.

\bibitem{chen2022large}
Z.~Chen, S.~Chen, Y.~Wu, Y.~Qian, C.~Wang, S.~Liu, Y.~Qian, and M.~Zeng,
  ``Large-scale self-supervised speech representation learning for automatic
  speaker verification,'' in \emph{Proc. ICASSP}.\hskip 1em plus 0.5em minus
  0.4em\relax IEEE, 2022, pp. 6147--6151.

\bibitem{NEURIPS2021_87682805}
H.-S. Choi, J.~Lee, W.~Kim, J.~Lee, H.~Heo, and K.~Lee, ``Neural analysis and
  synthesis: Reconstructing speech from self-supervised representations,'' in
  \emph{Proc. NIPS}, vol.~34, 2021, pp. 16\,251--16\,265.

\bibitem{9746430}
W.-C. Huang, S.-W. Yang, T.~Hayashi, H.-Y. Lee, S.~Watanabe, and T.~Toda,
  ``{S3PRL-VC}: Open-source voice conversion framework with self-supervised
  speech representations,'' in \emph{Proc. ICASSP}, 2022, pp. 6552--6556.

\bibitem{wang2023neural}
C.~Wang, S.~Chen, Y.~Wu, Z.~Zhang, L.~Zhou, S.~Liu, Z.~Chen, Y.~Liu, H.~Wang,
  J.~Li \emph{et~al.}, ``Neural codec language models are zero-shot text to
  speech synthesizers,'' \emph{arXiv preprint arXiv:2301.02111}, 2023.

\bibitem{liu23d_interspeech}
S.~Liu, Y.~Guo, C.~Du, X.~Chen, and K.~Yu, ``{DSE-TTS}: Dual speaker embedding
  for cross-lingual text-to-speech,'' in \emph{Proc. Interspeech}, 2023, pp.
  616--620.

\bibitem{9688093}
A.~Pasad, J.-C. Chou, and K.~Livescu, ``Layer-wise analysis of a
  self-supervised speech representation model,'' in \emph{Proc. ASRU}, 2021,
  pp. 914--921.

\bibitem{borsos2023audiolm}
Z.~Borsos, R.~Marinier, D.~Vincent, E.~Kharitonov, O.~Pietquin, M.~Sharifi,
  D.~Roblek, O.~Teboul, D.~Grangier, M.~Tagliasacchi \emph{et~al.}, ``{Audiolm:
  a language modeling approach to audio generation},'' \emph{IEEE/ACM
  Transactions on Audio, Speech, and Language Processing}, 2023.

\bibitem{kharitonov2023speak}
E.~Kharitonov, D.~Vincent, Z.~Borsos, R.~Marinier, S.~Girgin, O.~Pietquin,
  M.~Sharifi, M.~Tagliasacchi, and N.~Zeghidour, ``Speak, read and prompt:
  High-fidelity text-to-speech with minimal supervision,'' \emph{Transactions
  of the Association for Computational Linguistics}, vol.~11, pp. 1703--1718,
  2023.

\bibitem{le2024voicebox}
M.~Le, A.~Vyas, B.~Shi, B.~Karrer, L.~Sari, R.~Moritz, M.~Williamson,
  V.~Manohar, Y.~Adi, J.~Mahadeokar \emph{et~al.}, ``Voicebox: Text-guided
  multilingual universal speech generation at scale,'' \emph{Advances in neural
  information processing systems}, vol.~36, 2024.

\bibitem{zeghidour2021soundstream}
N.~Zeghidour, A.~Luebs, A.~Omran, J.~Skoglund, and M.~Tagliasacchi,
  ``Soundstream: An end-to-end neural audio codec,'' \emph{IEEE/ACM
  Transactions on Audio, Speech, and Language Processing}, vol.~30, pp.
  495--507, 2021.

\bibitem{defossez2023high}
A.~D{\'e}fossez, J.~Copet, G.~Synnaeve, and Y.~Adi, ``High fidelity neural
  audio compression,'' \emph{Transactions on Machine Learning Research}, 2023.

\bibitem{shen2023naturalspeech}
K.~Shen, Z.~Ju, X.~Tan, Y.~Liu, Y.~Leng, L.~He, T.~Qin, S.~Zhao, and J.~Bian,
  ``Naturalspeech 2: Latent diffusion models are natural and zero-shot speech
  and singing synthesizers,'' \emph{arXiv preprint arXiv:2304.09116}, 2023.

\bibitem{ju2024naturalspeech}
Z.~Ju, Y.~Wang, K.~Shen, X.~Tan, D.~Xin, D.~Yang, Y.~Liu, Y.~Leng, K.~Song,
  S.~Tang \emph{et~al.}, ``Naturalspeech 3: Zero-shot speech synthesis with
  factorized codec and diffusion models,'' \emph{arXiv preprint
  arXiv:2403.03100}, 2024.

\bibitem{lee2023hierspeech++}
S.-H. Lee, H.-Y. Choi, S.-B. Kim, and S.-W. Lee, ``Hierspeech++: Bridging the
  gap between semantic and acoustic representation of speech by hierarchical
  variational inference for zero-shot speech synthesis,'' \emph{arXiv preprint
  arXiv:2311.12454}, 2023.

\bibitem{arik2018neural}
S.~Arik, J.~Chen, K.~Peng, W.~Ping, and Y.~Zhou, ``Neural voice cloning with a
  few samples,'' in \emph{Proc. NIPS}, vol.~31, 2018.

\bibitem{wang2018style}
Y.~Wang, D.~Stanton, Y.~Zhang, R.-S. Ryan, E.~Battenberg, J.~Shor, Y.~Xiao,
  Y.~Jia, F.~Ren, and R.~A. Saurous, ``Style tokens: Unsupervised style
  modeling, control and transfer in end-to-end speech synthesis,'' in
  \emph{Proc. ICML}.\hskip 1em plus 0.5em minus 0.4em\relax PMLR, 2018, pp.
  5180--5189.

\bibitem{casanova2021sc}
E.~Casanova, C.~Shulby, E.~G{\"o}lge, N.~M. M{\"u}ller, F.~S. {de Oliveira},
  A.~Candido~Jr., A.~{da Silva Soares}, S.~M. Aluisio, and M.~A. Ponti,
  ``{SC-GlowTTS}: An efficient zero-shot multi-speaker text-to-speech model,''
  in \emph{Proc. {{Interspeech}}}, 2021, pp. 3645--3649.

\bibitem{xin2021disentangled}
D.~Xin, T.~Komatsu, S.~Takamichi, and H.~Saruwatari, ``Disentangled speaker and
  language representations using mutual information minimization and domain
  adaptation for cross-lingual {TTS},'' in \emph{Proc. ICASSP}.\hskip 1em plus
  0.5em minus 0.4em\relax IEEE, 2021, pp. 6608--6612.

\bibitem{yang2022cross}
J.~Yang and L.~He, ``Cross-lingual text-to-speech using multi-task learning and
  speaker classifier joint training,'' \emph{arXiv preprint arXiv:2201.08124},
  2022.

\bibitem{DBLP:conf/aaai/Li0LZL19}
N.~Li, S.~Liu, Y.~Liu, S.~Zhao, and M.~Liu, ``Neural speech synthesis with
  {Transformer} network,'' in \emph{Proc. AAAI}, 2019, pp. 6706--6713.

\bibitem{NEURIPS2019_f63f65b5}
Y.~Ren, Y.~Ruan, X.~Tan, T.~Qin, S.~Zhao, Z.~Zhao, and T.-Y. Liu, ``Fastspeech:
  Fast, robust and controllable text to speech,'' in \emph{Advances in Neural
  Information Processing Systems}, vol.~32, 2019.

\bibitem{DBLP:conf/icassp/Lancucki21}
A.~Lancucki, ``{FastPitch}: Parallel text-to-speech with pitch prediction,'' in
  \emph{Proc. ICASSP}.\hskip 1em plus 0.5em minus 0.4em\relax {IEEE}, 2021, pp.
  6588--6592.

\bibitem{yang20g_interspeech}
J.~Yang and L.~He, ``Towards universal text-to-speech,'' in \emph{Proc.
  Interspeech}, 2020, pp. 3171--3175.

\bibitem{DBLP:conf/ssw/OordDZSVGKSK16}
A.~van~den Oord, S.~Dieleman, H.~Zen, K.~Simonyan, O.~Vinyals, A.~Graves,
  N.~Kalchbrenner, A.~W. Senior, and K.~Kavukcuoglu, ``Wavenet: {A} generative
  model for raw audio,'' in \emph{ISCA, September 2016}.\hskip 1em plus 0.5em
  minus 0.4em\relax {ISCA}, 2016, p. 125.

\bibitem{badlani23_interspeech}
R.~Badlani, R.~Valle, K.~J. Shih, J.~F. Santos, S.~Gururani, and B.~Catanzaro,
  ``{RAD-MMM}: Multilingual multiaccented multispeaker text to speech,'' in
  \emph{Proc. Interspeech}, 2023, pp. 626--630.

\bibitem{shih2021rad}
K.~J. Shih, R.~Valle, R.~Badlani, A.~Lancucki, W.~Ping, and B.~Catanzaro,
  ``{RAD-TTS}: Parallel flow-based {TTS} with robust alignment learning and
  diverse synthesis,'' in \emph{Proc. ICML Workshop on Invertible Neural
  Networks, Normalizing Flows, and Explicit Likelihood Models}, 2021.

\bibitem{DBLP:conf/interspeech/ChoJLW22}
H.~Cho, W.~Jung, J.~Lee, and S.~H. Woo, ``{SANE-TTS}: Stable and natural
  end-to-end multilingual text-to-speech,'' in \emph{Proc. Interspeech}.\hskip
  1em plus 0.5em minus 0.4em\relax {ISCA}, 2022.

\bibitem{kim2021conditional}
J.~Kim, J.~Kong, and J.~Son, ``{Conditional variational autoencoder with
  adversarial learning for end-to-end text-to-speech},'' in \emph{Proc.
  ICML}.\hskip 1em plus 0.5em minus 0.4em\relax PMLR, 2021, pp. 5530--5540.

\bibitem{DBLP:conf/interspeech/ChenCLMCWX19}
M.~Chen, M.~Chen, S.~Liang, J.~Ma, L.~Chen, S.~Wang, and J.~Xiao,
  ``Cross-lingual, multi-speaker text-to-speech synthesis using neural speaker
  embedding,'' in \emph{Proc. Interspeech}.\hskip 1em plus 0.5em minus
  0.4em\relax {ISCA}, 2019, pp. 2105--2109.

\bibitem{10444075}
T.~Saeki, S.~Maiti, X.~Li, S.~Watanabe, S.~Takamichi, and H.~Saruwatari,
  ``Text-inductive graphone-based language adaptation for low-resource speech
  synthesis,'' \emph{IEEE/ACM Transactions on Audio, Speech, and Language
  Processing}, vol.~32, pp. 1829--1844, 2024.

\bibitem{DBLP:conf/icassp/LiZSWC19}
B.~Li, Y.~Zhang, T.~N. Sainath, Y.~Wu, and W.~Chan, ``Bytes are all you need:
  End-to-end multilingual speech recognition and synthesis with bytes,'' in
  \emph{Proc. ICASSP}.\hskip 1em plus 0.5em minus 0.4em\relax {IEEE}, 2019, pp.
  5621--5625.

\bibitem{staib20_interspeech}
M.~Staib, T.~H. Teh, A.~Torresquintero, D.~S.~R. Mohan, L.~Foglianti,
  R.~Lenain, and J.~Gao, ``{Phonological Features for 0-Shot Multilingual
  Speech Synthesis},'' in \emph{Proc. Interspeech 2020}, 2020, pp. 2942--2946.

\bibitem{wells2021cross}
D.~Wells and K.~Richmond, ``{Cross-lingual transfer of phonological features
  for low-resource speech synthesis},'' in \emph{Proc. SSW}, 2021, pp.
  160--165.

\bibitem{Mortensen-et-al:2018}
D.~R. Mortensen, S.~Dalmia, and P.~Littell,
  ``\BIBforeignlanguage{english}{{Epitran}: Precision {G2P} for many
  languages},'' in \emph{\BIBforeignlanguage{english}{Proc. LREC}}, Paris,
  France, May 2018.

\bibitem{wang2022wav2vec}
Y.~Wang, J.~Li, H.~Wang, Y.~Qian, C.~Wang, and Y.~Wu, ``Wav2vec-switch:
  Contrastive learning from original-noisy speech pairs for robust speech
  recognition,'' in \emph{Proc. ICASSP}.\hskip 1em plus 0.5em minus 0.4em\relax
  IEEE, 2022, pp. 7097--7101.

\bibitem{DBLP:conf/interspeech/SiuzdakDRJ22}
H.~Siuzdak, P.~Dura, P.~van Rijn, and N.~Jacoby, ``{WavThruVec}: Latent speech
  representation as intermediate features for neural speech synthesis,'' in
  \emph{Proc. Interspeech}.\hskip 1em plus 0.5em minus 0.4em\relax {ISCA},
  2022, pp. 833--837.

\bibitem{lin2021s2vc}
J.-h. Lin, Y.~Y. Lin, C.-M. Chien, and H.-y. Lee, ``{{S2VC}}: {{A}} framework
  for any-to-any voice conversion with self-supervised pretrained
  representations,'' in \emph{Proc. {{Interspeech}}}, 2021, pp. 836--840.

\bibitem{chen2022does}
S.~Chen, Y.~Wu, C.~Wang, S.~Liu, Z.~Chen, P.~Wang, G.~Liu, J.~Li, J.~Wu, X.~Yu,
  and F.~Wei, ``Why does {{Self-Supervised Learning}} for {{Speech Recognition
  Benefit Speaker Recognition}}?'' in \emph{Proc. Interspeech}.\hskip 1em plus
  0.5em minus 0.4em\relax {ISCA}, Sep. 2022, pp. 3699--3703.

\bibitem{10129796}
Y.-J. Zhang, C.~Zhang, W.~Song, Z.~Zhang, Y.~Wu, and X.~He, ``Prosody modelling
  with pre-trained cross-utterance representations for improved speech
  synthesis,'' \emph{IEEE/ACM Transactions on Audio, Speech, and Language
  Processing}, vol.~31, pp. 2812--2823, 2023.

\bibitem{DBLP:journals/corr/abs-2006-13979}
A.~Conneau, A.~Baevski, R.~Collobert, A.~Mohamed, and M.~Auli, ``Unsupervised
  cross-lingual representation learning for speech recognition,'' in
  \emph{Proc. Interspeech}.\hskip 1em plus 0.5em minus 0.4em\relax {ISCA}, Aug.
  2021, pp. 2426--2430.

\bibitem{du2022vqtts}
C.~Du, Y.~Guo, X.~Chen, and K.~Yu, ``{VQTTS: High-Fidelity Text-to-Speech
  Synthesis with Self-Supervised VQ Acoustic Feature},'' in \emph{Proc.
  Interspeech}, {Incheon, Korea}, {Sep.} 2022, pp. 1596--1600.

\bibitem{chen2023vector}
L.-W. Chen, S.~Watanabe, and A.~Rudnicky, ``A vector quantized approach for
  text to speech synthesis on real-world spontaneous speech,'' in
  \emph{Proceedings of the AAAI Conference on Artificial Intelligence},
  vol.~37, no.~11, 2023, pp. 12\,644--12\,652.

\bibitem{liu22p_interspeech}
C.~Liu, Z.-H. Ling, and L.-H. Chen, ``Pronunciation dictionary-free
  multilingual speech synthesis by combining unsupervised and supervised
  phonetic representations,'' in \emph{Proc. Interspeech}, 2022, pp.
  4282--4286.

\bibitem{wells23_interspeech}
D.~Wells, K.~Richmond, and W.~Lamb, ``{A Low-Resource Pipeline for
  Text-to-Speech from Found Data With Application to Scottish Gaelic},'' in
  \emph{Proc. INTERSPEECH 2023}, 2023, pp. 4324--4328.

\bibitem{saeki2024extending}
T.~Saeki, G.~Wang, N.~Morioka, I.~Elias, K.~Kastner, A.~Rosenberg,
  B.~Ramabhadran, H.~Zen, F.~Beaufays, and H.~Shemtov, ``Extending multilingual
  speech synthesis to 100+ languages without transcribed data,'' \emph{arXiv
  preprint arXiv:2402.18932}, 2024.

\bibitem{lee2022hierspeech}
S.-H. Lee, S.-B. Kim, J.-H. Lee, E.~Song, M.-J. Hwang, and S.-W. Lee,
  ``Hierspeech: Bridging the gap between text and speech by hierarchical
  variational inference using self-supervised representations for speech
  synthesis,'' \emph{Advances in Neural Information Processing Systems},
  vol.~35, pp. 16\,624--16\,636, 2022.

\bibitem{li2024styletts}
Y.~A. Li, C.~Han, V.~Raghavan, G.~Mischler, and N.~Mesgarani, ``Styletts 2:
  Towards human-level text-to-speech through style diffusion and adversarial
  training with large speech language models,'' \emph{Advances in Neural
  Information Processing Systems}, vol.~36, 2024.

\bibitem{zhang2023speak}
Z.~Zhang, L.~Zhou, C.~Wang, S.~Chen, Y.~Wu, S.~Liu, Z.~Chen, Y.~Liu, H.~Wang,
  J.~Li \emph{et~al.}, ``Speak foreign languages with your own voice:
  Cross-lingual neural codec language modeling,'' \emph{arXiv preprint
  arXiv:2303.03926}, 2023.

\bibitem{barrault2023seamless}
L.~Barrault, Y.-A. Chung, M.~C. Meglioli, D.~Dale, N.~Dong, M.~Duppenthaler,
  P.-A. Duquenne, B.~Ellis, H.~Elsahar, J.~Haaheim \emph{et~al.}, ``Seamless:
  Multilingual expressive and streaming speech translation,'' \emph{arXiv
  preprint arXiv:2312.05187}, 2023.

\bibitem{NEURIPS2020_c5d73680}
J.~Kong, J.~Kim, and J.~Bae, ``{HiFi-GAN}: Generative adversarial networks for
  efficient and high fidelity speech synthesis,'' in \emph{Proc. NIPS},
  vol.~33.\hskip 1em plus 0.5em minus 0.4em\relax Curran Associates, Inc.,
  2020, pp. 17\,022--17\,033.

\bibitem{badlani2022one}
R.~Badlani, A.~{\L}a{\'n}cucki, K.~J. Shih, R.~Valle, W.~Ping, and
  B.~Catanzaro, ``One {TTS} alignment to rule them all,'' in \emph{Proc.
  ICASSP}.\hskip 1em plus 0.5em minus 0.4em\relax IEEE, 2022, pp. 6092--6096.

\bibitem{berrebbi2022combining}
D.~Berrebbi, J.~Shi, B.~Yan, O.~López-Francisco, J.~Amith, and S.~Watanabe,
  ``Combining spectral and self-supervised features for low resource speech
  recognition and translation,'' in \emph{Proc. Interspeech}, 2022, pp.
  3533--3537.

\bibitem{guo2022multi}
H.~Guo, F.~Xie, F.~K. Soong, X.~Wu, and H.~Meng, ``A multi-stage multi-codebook
  {VQ-VAE} approach to high-performance neural {TTS},'' in \emph{Proc.
  Interspeech}.\hskip 1em plus 0.5em minus 0.4em\relax {ISCA}, 2022, pp.
  1611--1615.

\bibitem{guo2022towards}
H.~Guo, F.~Xie, X.~Wu, F.~K. Soong, and H.~Meng, ``{MSMC-TTS}: Multi-stage
  multi-codebook {VQ-VAE} based neural {TTS},'' \emph{IEEE/ACM Transactions on
  Audio, Speech, and Language Processing}, vol.~31, pp. 1811--1824, 2023.

\bibitem{jang2021univnet}
W.~Jang, D.~Lim, J.~Yoon, B.~Kim, and J.~Kim, ``{UnivNet: A neural vocoder with
  multi-resolution spectrogram discriminators for high-fidelity waveform
  generation},'' in \emph{Proc. INTERSPEECH}, 2021, pp. 2207--2211.

\bibitem{Pratap2020MLSAL}
V.~Pratap, Q.~Xu, A.~Sriram, G.~Synnaeve, and R.~Collobert, ``{MLS}: A
  large-scale multilingual dataset for speech research,'' in \emph{Proc.
  Interspeech}, 2020, p. 2757–2761.

\bibitem{6639248}
T.~Schultz, N.~T. Vu, and T.~Schlippe, ``{GlobalPhone}: A multilingual text \&
  speech database in 20 languages,'' in \emph{Proc. ICASSP}, 2013, pp.
  8126--8130.

\bibitem{park2019css10}
K.~Park and T.~Mulc, ``{CSS10}: A collection of single speaker speech datasets
  for 10 languages,'' in \emph{Proc. Interspeech}, 2019, pp. 1566--1570.

\bibitem{ljspeech17}
K.~Ito and L.~Johnson, ``The {LJ} speech dataset,''
  \url{https://keithito.com/LJ-Speech-Dataset/}, 2017.

\bibitem{NST}
N.~L. Technology, ``{NST} {Swedish} speech synthesis,''
  \url{https://www.nb.no/sprakbanken/en/resource-catalogue/oai-nb-no-sbr-18/},
  2003.

\bibitem{sv56}
\BIBentryALTinterwordspacing
I.~T. Union. {{Recommendation G.191}: Software Tools and Audio Coding
  Standardization}. (2005, Nov 11). [Online]. Available:
  \url{https://www.itu.int/rec/T-REC-P.56/en}
\BIBentrySTDinterwordspacing

\bibitem{heo2020clova}
H.~S. Heo, B.-J. Lee, J.~Huh, and J.~S. Chung, ``{CLOVA} baseline system for
  the {VoxCeleb Speaker Recognition Challenge 2020},'' \emph{arXiv preprint
  arXiv:2009.14153}, 2020.

\bibitem{biaobei}
\BIBentryALTinterwordspacing
{{Data-baker}}. {{Chinese} Standard {Mandarin} Speech Copus}. (2022, Nov).
  [Online]. Available: \url{https://www.data-baker.com/open_source.html}
\BIBentrySTDinterwordspacing

\bibitem{van2008visualizing}
L.~Van~der Maaten and G.~Hinton, ``{Visualizing data using t-SNE.}''
  \emph{Journal of machine learning research}, vol.~9, no.~11, 2008.

\bibitem{do2022text}
P.~Do, M.~Coler, J.~Dijkstra, and E.~Klabbers, ``Text-to-speech for
  under-resourced languages: Phoneme mapping and source language selection in
  transfer learning,'' in \emph{Proc. ELRA/ISCA SIG on Under-Resourced
  Languages}, 2022, pp. 16--22.

\bibitem{saeki2022utmos}
T.~Saeki, D.~Xin, W.~Nakata, T.~Koriyama, S.~Takamichi, and H.~Saruwatari,
  ``{UTMOS: UTokyo-SaruLab system for VoiceMOS Challenge 2022},'' in
  \emph{Proc. Interspeech}, 2022, pp. 4521--4525.

\bibitem{zen2019libritts}
H.~Zen, V.~Dang, R.~Clark, Y.~Zhang, R.~J. Weiss, Y.~Jia, Z.~Chen, and Y.~Wu,
  ``{Libritts: A corpus derived from librispeech for text-to-speech},''
  \emph{arXiv preprint arXiv:1904.02882}, 2019.

\bibitem{panayotov2015librispeech}
V.~Panayotov, G.~Chen, D.~Povey, and S.~Khudanpur, ``{Librispeech: an asr
  corpus based on public domain audio books},'' in \emph{2015 IEEE
  international conference on acoustics, speech and signal processing
  (ICASSP)}.\hskip 1em plus 0.5em minus 0.4em\relax IEEE, 2015, pp. 5206--5210.

\end{thebibliography}

\newpage
\begin{IEEEbiography}[{\includegraphics[width=1in,height=1.25in,clip,keepaspectratio]{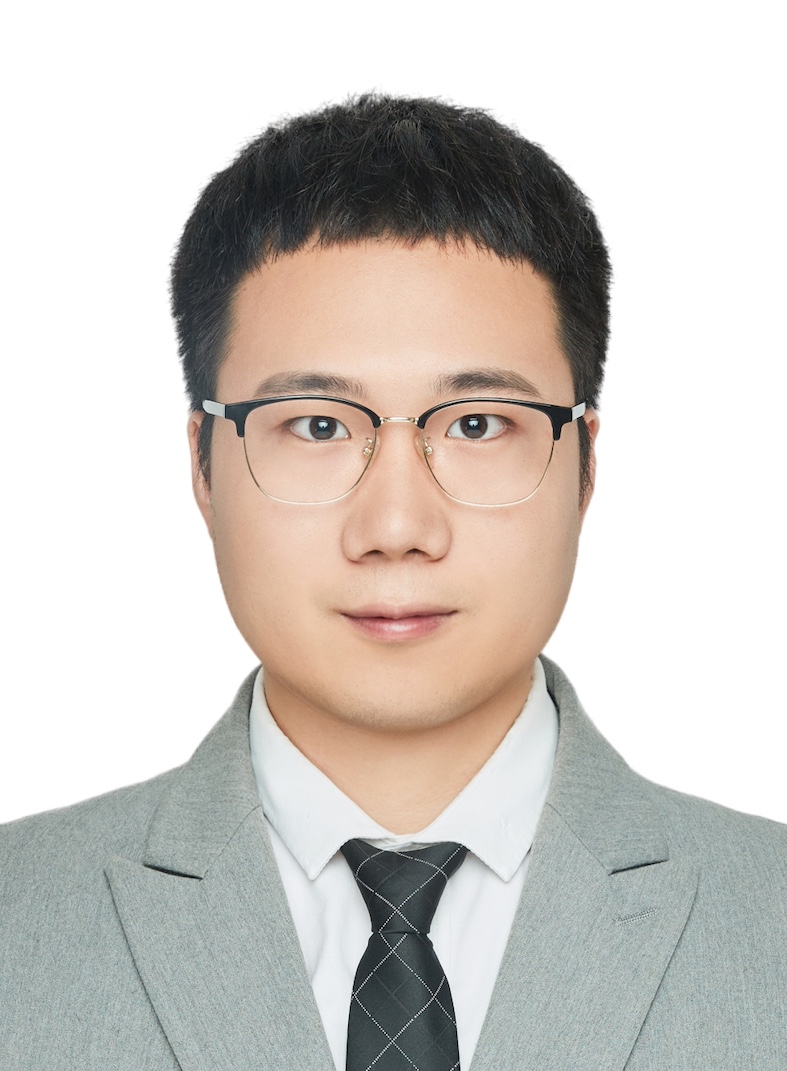}}]{Cheng Gong} received his B.S. and  M.S. degree from the Hohai University, Nanjing, China, in 2016 and 2019. He is currently working toward the Ph.D. degree in Tianjin Key Laboratory of Cognitive Computing and Application, Tianjin University, Tianjin, China. He is also an exchange Ph.D. candidate at the National Institute of Informatics, Japan, funded by China Scholarship Council (CSC). His research interests include expressive speech synthesis and multilingual speech synthesis. 

\end{IEEEbiography}

\begin{IEEEbiography}[{\includegraphics[width=1in,height=1.25in,clip,keepaspectratio]{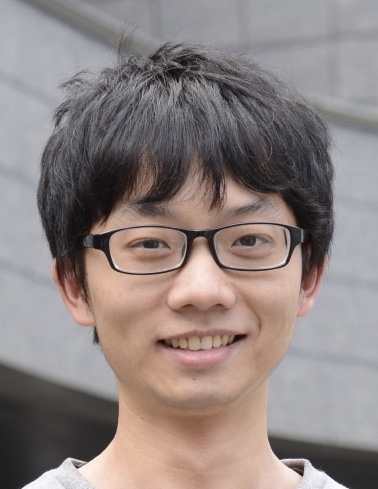}}]{Xin Wang} is a project associate professor at National Institute of Informatics and a JST PRESTO researcher, Japan. He was among the organizing team of the recent ASVspoof Challenges and the VoicePrivacy initiatives in 2020 and 2022. His research focuses on speech audio generation, text-to-speech synthesis, anti-spoofing, and other speech security and private related tasks. He is a Member of IEEE.
\end{IEEEbiography}
\begin{IEEEbiography}[{\includegraphics[width=1in,height=1.25in,clip,keepaspectratio]{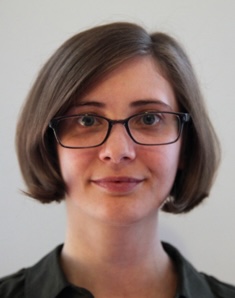}}]{Erica Cooper} received the B.Sc. and M.Eng. degrees in electrical engineering and computer science from the Massachusetts Institute of Technology, Cambridge, MA, USA, in 2009 and 2010, respectively, and the Ph.D. degree in computer science from Columbia University, New York, NY, USA, in 2019. She was a postdoctoral researcher at the National Institute of Informatics, Tokyo, Japan from 2019 to 2024. Her research interests include statistical machine learning, speech synthesis, and speech quality assessment. 
\end{IEEEbiography} 
\begin{IEEEbiography}[{\includegraphics[width=1in,height=1.25in,clip,keepaspectratio]{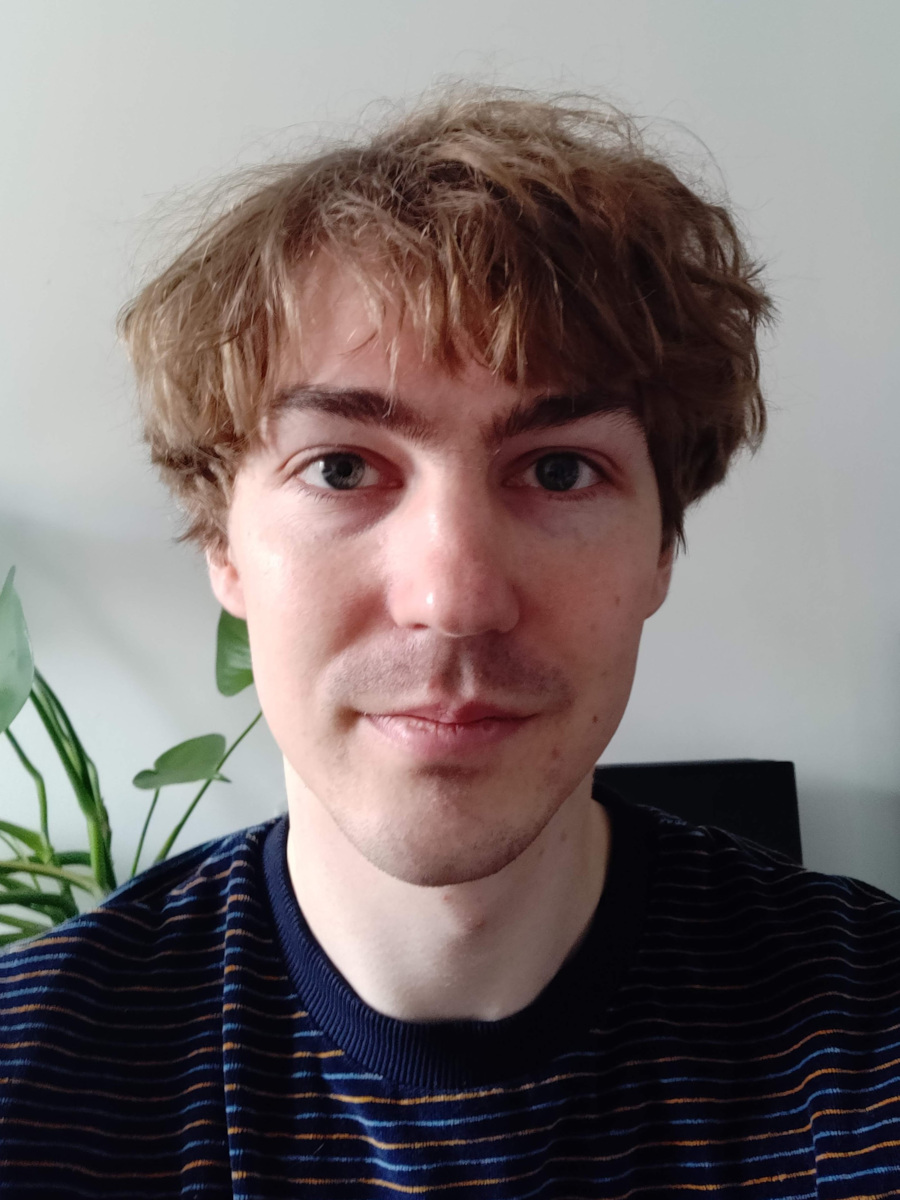}}]{Dan Wells} is a PhD student at the Centre for Speech Technology Research, University of Edinburgh. He received the BA degree in Linguistics from the University of Cambridge in 2013 and the MSc in Speech and Language Processing from the University of Edinburgh in 2015. His research interests are in speech synthesis, in particular methods for reducing linguistic data requirements when building a voice for a new language, and speech representation learning. He also has several years of industry experience in automatic speaker and speech recognition. 
\end{IEEEbiography}

\begin{IEEEbiography}[{\includegraphics[width=1in,height=1.25in,clip,keepaspectratio]{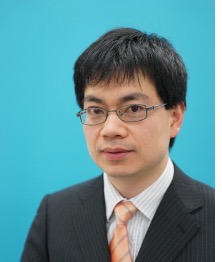}}]{Longbiao Wang}
 received his Dr. Eng. degree from Toyohashi University of Technology, Japan, in 2008. He was an assistant professor in the faculty of Engineering at Shizuoka University, Japan from April 2008 to September 2012. He was an associate professor at Nagaoka University of Technology, Japan from Oct. 2012 to Aug. 2016. He is currently a professor, director of Tianjin Key Laboratory of Cognitive Computing and Application and vice dean of School of Artificial Intelligence at Tianjin University, China. His research interests include robust speech recognition, speaker recognition, acoustic signal processing and natural language processing. 
\end{IEEEbiography}

\begin{IEEEbiography}[{\includegraphics[width=1in,height=1.25in,clip,keepaspectratio]{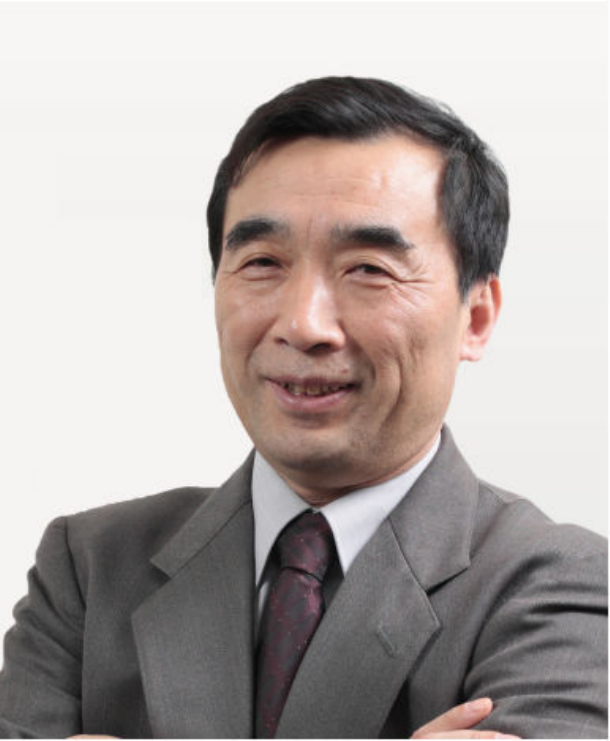}}]{Jianwu Dang}
 (M’12) graduated from Tsinghua Univ., China, in 1982, and got his M.S. degree at the same university in 1984. He worked for Tianjin Univ. as a lecture from 1984 to 1988. He was awarded the PhD degree from Shizuoka Univ., Japan in 1992. He worked for ATR Human Information
 Processing Labs., Japan, as a senior researcher from
 1992 to 2001. He joined the University of Waterloo,
 Canada, as a visiting scholar for one year from 1998.
 Since 2001, he has worked for Japan Advanced
 Institute of Science and Technology (JAIST) as a
 professor. He joined the Institute of Communication Parlee (ICP), Center of
 National Research Scientific, France, as a research scientist the first class from
 2002 to 2003. Since 2009, he has joined Tianjin University, Tianjin, China.
 His research interests are in all the fields of speech science including brain
 science, and speech signal processing. He built MRI-based bio-physiological
 models for speech and swallowing, and endeavors to apply these models on
 clinics.
\end{IEEEbiography}

\begin{IEEEbiography}[{\includegraphics[width=1in,height=1.25in,clip,keepaspectratio]{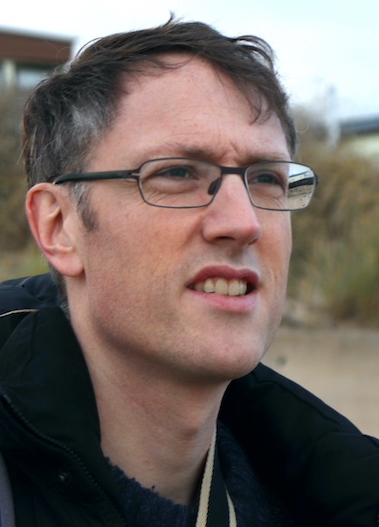}}]{Korin Richmond}
is a Reader in Speech Technology (UK Associate Professor) at the Centre for Speech Technology Research, University of Edinburgh.  With experience in human language and speech technology going back to 1991, he received an MA (Linguistics and Russian, 1995), MSc (Cognitive Science and Natural Language Processing, 1997) and PhD (titled “Estimating Articulatory Parameters from the Acoustic Speech Signal”, 2002) from the University of Edinburgh.  He has lectured at the University of Edinburgh since 2016, and has diverse research interests, including: speech synthesis; pronunciation modelling and lexicography; speech therapy; articulatory data and modelling.  He has over 110 publications in speech technology and processing (h-index 33; i10-index 66).  He is an IEEE Senior Fellow and has served two 3-year terms as a member of the IEEE Speech and Language Technical Committee. 
\end{IEEEbiography}

\begin{IEEEbiography}[{\includegraphics[width=1in,height=1.25in,clip,keepaspectratio]{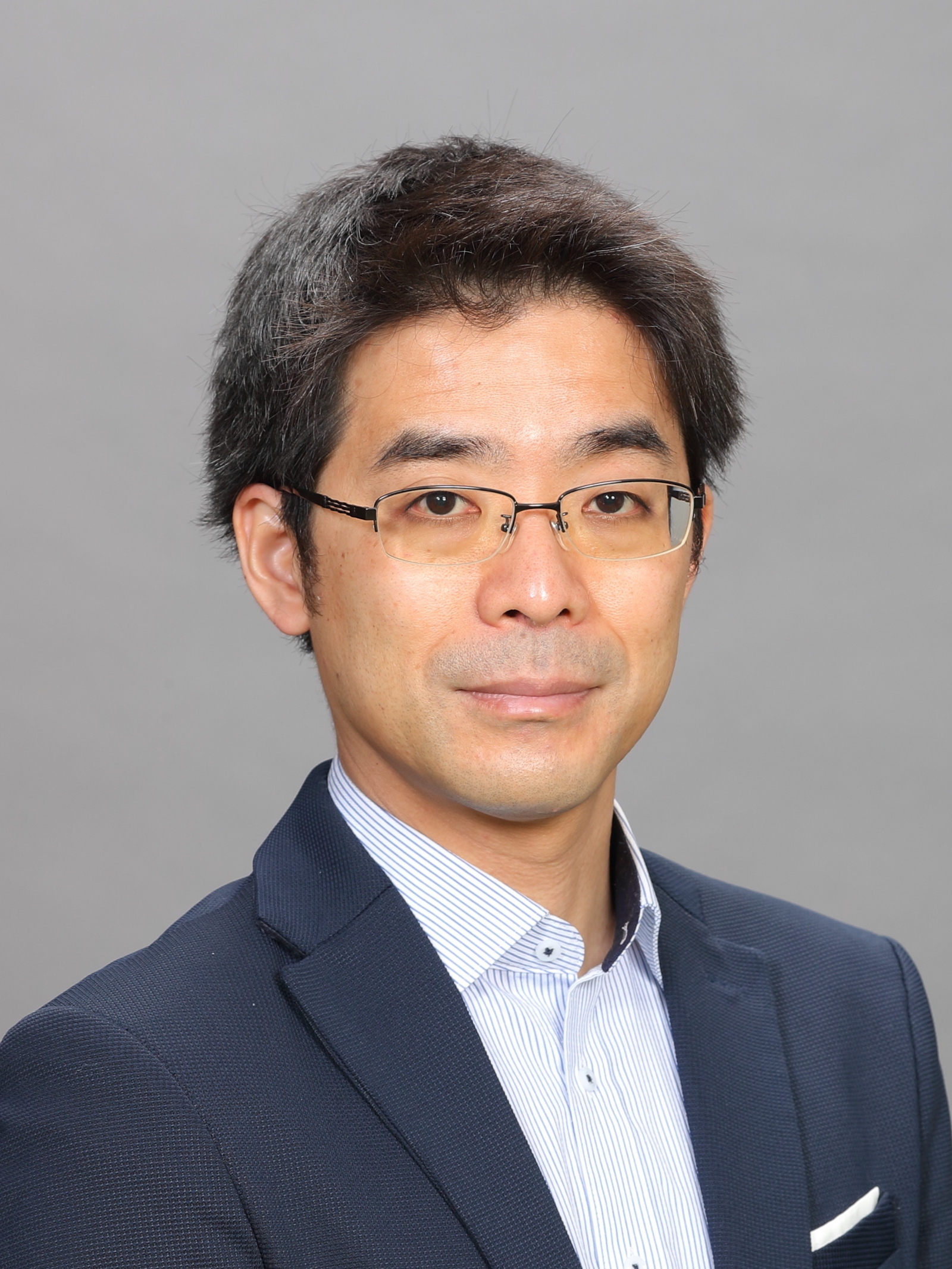}}]{Junichi Yamagishi} (Senior Member, IEEE) received a Ph.D. degree from the Tokyo Institute of Technology (Tokyo Tech), Tokyo, Japan, in 2006. From 2007 to 2013 he was a research fellow in the Centre for Speech Technology Research, University of Edinburgh, U.K. He became an associate professor with the National Institute of Informatics, Japan, in 2013, where he is currently a professor. His research interests include speech processing, machine learning, signal processing, biometrics, digital media cloning, and media forensics. 
\end{IEEEbiography}
\end{document}